\documentclass[twocolumn,amsmath,amssymb,prc,superscriptaddress,floatfix,showpacs,nofootinbib]{revtex4-1}

\usepackage{amsmath}
\usepackage{amssymb}
\usepackage{epsf}
\usepackage{epsfig}
\usepackage{graphicx}
\usepackage[english]{babel}
\usepackage[latin1]{inputenc}
\usepackage{rotating}
\usepackage{subfigure}
\usepackage{color}
\usepackage{slashed}
\usepackage{wrapfig}

\usepackage{txfonts}

\newcommand{\bea}{\begin{eqnarray}}
\newcommand{\eea}{\end{eqnarray}}

\newcommand{\mywidth}{8.5cm}

\begin{document}

\title{Amplified Fermion Production from Overpopulated Bose Fields}

\author{J. Berges}
\email[]{j.berges@thphys.uni-heidelberg.de}
\affiliation{Institut f\"{u}r Theoretische Physik, Universit\"{a}t Heidelberg,
  Philosophenweg 16, 69120 Heidelberg, Germany}
\affiliation{ ExtreMe Matter Institute EMMI, 
Planckstra\ss e~1, 64291~Darmstadt, Germany}

\author{D. Gelfand}
\email[]{d.gelfand@thphys.uni-heidelberg.de}
\affiliation{Institut f\"{u}r Theoretische Physik, Universit\"{a}t Heidelberg,
  Philosophenweg 16, 69120 Heidelberg, Germany}
  
\author{D. Sexty}
\email[]{d.sexty@thphys.uni-heidelberg.de}
\affiliation{Institut f\"{u}r Theoretische Physik, Universit\"{a}t Heidelberg,
  Philosophenweg 16, 69120 Heidelberg, Germany}

\begin{abstract}
We study the real-time dynamics of fermions coupled to scalar fields  
in a linear sigma model, which is often employed in the context of  
preheating after inflation or as a low-energy effective model for  
quantum chromodynamics. We find a dramatic amplification of  
fermion production in the presence of highly occupied bosonic quanta  
for weak as well as strong couplings. For this we consider the range  
of validity of different methods: lattice simulations with male/female  
fermions, the mode functions approach and the quantum 2PI effective  
action with its associated kinetic theory. For strongly coupled  
fermions we find a rapid approach to a Fermi-Dirac distribution with  
time-dependent temperature and chemical potential parameters, while  
the bosons are still far from equilibrium.
\end{abstract}
\maketitle

\section{Introduction}\label{Introduction}
Understanding the real-time dynamics of fermions coupled to highly ocupied bosonic fields
 is central for a wide range  
of physical systems. Important examples concern the creation of matter  
fields from inflaton decay during preheating in inflationary  
cosmology, the fermion production from gauge fields during the early  
stages of ultra-relativistic collision experiments of heavy nuclei, or  
non-relativistic systems of ultracold quantum gases with fermions and  
bosons.

Though these examples cover a vast range of characteristic energy  
scales, they require rather similar theoretical descriptions. Standard  
semi-classical approaches are based on solutions of a Dirac equation,  
or its non-relativistic limit, in the presence of a time-dependent  
bosonic background field. In this context, large coherent Bose fields  
are often treated classically. However, since identical fermions  
cannot occupy the same state, their quantum nature is highly relevant  
and a consistent quantum treatment of the fermion dynamics is of  
crucial importance. It has been pointed out that semi-classical  
approaches can fail to describe the fermion dynamics in the presence  
of high occupancies of bosonic quanta~\cite{BergesGelfandPruschke}.  
For the example of a decaying inflaton field, it was shown that  
quantum corrections dramatically enhance the production of fermions  
following preheating in the early Universe.

In this work, we analyze the real-time dynamics of Dirac fermions  
coupled to scalar fields using different  
methods. We consider a linear sigma model, which is often employed in  
the context of preheating after inflation or also as a low-energy  
effective model for quantum chromodynamics. We compare different  
real-time techniques: lattice simulations with male/female  
fermions~\cite{BorsanyiHindmarsh}, the mode functions  
approach~\cite{AartsSmit} and the quantum 2PI effective  
action~\cite{Berges:2002wr} with its associated kinetic theory.
We discuss their range of applicability in detail. As our main result,  
we confirm the dramatic amplification of fermion production in the  
presence of highly occupied bosons that was first pointed out in  
Ref.~\cite{BergesGelfandPruschke} and extend the results to the strong  
coupling regime.

It turns out that the efficient male/female lattice approach  
accurately converges to the exact mode functions result for the  
available lattice sizes. Our study shows the strength of the  
male/female method to address physical questiones for large volumes~\cite{BergesGelfandPruschke,HebenstreitBergesGelfand,StringBreaking,SaffinTranberg1,SaffinTranberg2,MouSaffinTranberg},  
something where the mode function approach becomes computationally  
intractable. For weak couplings we find that the lattice simulation  
results agree well with those obtained from the quantum 2PI effective  
action, emphasizing the ability of the lattice approach to describe  
genuine quantum phenomena.

Applying an improved lattice discretization with a pseudoscalar Wilson  
term, we accurately resolve for the first time the high-momentum  
behavior of particle number distributions. For weak couplings this  
reveals a power-law behavior above a characteristic momentum. For  
strongly coupled fermions, we find a rapid approach to a quasi-thermal  
Fermi-Dirac distribution with time-dependent temperature and chemical  
potential parameters. Remarkably, this happens while the bosons are  
still showing turbulent behavior far from equilibrium.

The paper is organized as follows. In Sec. \ref{model} and \ref{method} we describe the  
model we use to simulate fermion production, explain the male/female  
method and mode functions approach. We also elaborate on details of  
our lattice formulation, our choice of initial conditions and  
renormalization procedure. In Sec. \ref{mode functions} we demonstrate  
the convergence of male/female method towards the exact mode functions  
results. In Sec. \ref{2PI} consequences of neglecting  
higher-order quantum fluctuations as well as details of our 2PI  
approach are discussed. Also in \ref{2PI} we present numerical  
evidence for applicability of 2PI and the degree of agreement between  
2PI and lattice simulations. In Sec. \ref{parametric} and \ref{kinetic} we finally  
arrive at our results for fermion production from parametric resonance  
at strong and weak coupling. We summarize and conclude in Sec.  
\ref{Conclusions}.

\section{Model and initial conditions}
\label{model}

We consider a relativistic quantum field theory of coupled bosonic 
and fermionic degrees of freedom. It describes a generic linear sigma model for a $N_s = 4$ component scalar field $( \sigma, \vec{\pi} )$ with self-coupling $\lambda$. The scalars interacts via a Yukawa coupling $g$ with $N_f =2$ flavors of massless Dirac fermions $\psi_i$ with flavor index $i$. The Lagrangian density is given by\footnote{Summation over repeated indices is implied, the metric tensor has signature $(+,-,-,-)$ and we use natural units with $\hbar=c=k_B=1$.} 
\begin{eqnarray}
\mathcal{L} &= &  \frac{1}{2} \left( \partial_{\mu}\sigma\, \partial^{\mu} \sigma + \partial_{\mu}\vec{\pi}\, \partial^{\mu} \vec{\pi} \right) - \frac{1}{2}\, m^{2} \left( \sigma^2 + \vec{\pi}^2 \right)  
\nonumber\\
&-&  \frac{\lambda}{4! N_s}\left(\sigma^2 + \vec{\pi}^2\right)^{2} + \bar{\psi}\left(i\,\partial_{\mu}\gamma^{\mu}\right)\psi - \frac{g}{N_f}\bar{\psi}\left(\sigma + i\gamma_{5} \vec{\tau} \vec{\pi} \right)\psi,
\nonumber\\
\label{eq:lagrangian}
\end{eqnarray}
with Dirac matrices $\gamma^\mu$ ($\mu = 0, \ldots, 3$), $\gamma^5= \gamma_5 = i \gamma^0\gamma^1 \gamma^2 \gamma^3$ and $\bar{\psi} \equiv \psi^\dagger \gamma^0$. We denote by $\vec{\tau}$ the Pauli matrices and space-time variables by $x\equiv (x^0,{\bf x})$. 

The equal-time anti-commutation relations for the fermions are encoded in the spectral function
\begin{equation}
\rho_{\psi,ij}(x,y) \equiv i \langle \{ \psi_i(x), \bar \psi_j(y) \} \rangle
\end{equation}
as
\begin{equation}
\gamma^0 \rho_{\psi,ij}(x,y)|_{x^0=y^0} = i \delta({\bf x}- {\bf y})\, \delta_{ij}
\end{equation}
with $\{ A, B \} \equiv AB + BA$. Correspondingly, the boson commutation relations are encoded in  
\begin{equation}
\rho_{\sigma}(x,y) \equiv i \langle [ \sigma(x), \sigma(y) ] \rangle
\end{equation}
with
\begin{equation}
\partial_{x^0} \rho_{\sigma}(x,y)|_{x^0=y^0} = \delta({\bf x}- {\bf y})
\end{equation}
for $[A,B] \equiv AB - BA$. Equivalently, one can define spectral functions for the $\vec{\pi}$ fields, with vanishing commutors between different fields. 

The brackets $\langle A \rangle \equiv {\rm tr}(\varrho_0 A)$ denote the trace over a normalized initial density matrix $\varrho_0$, which specifies the initial conditions at time $t_0$. Here we will choose Gaussian initial conditions, where $\varrho_0$ is completely determined by one- and two-point correlation functions at $t_0$. This class of initial conditions will allow us to study, in particular, particle production from nonequilibrium instabilities as will be discussed below. We emphasize that a choice of initial conditions does not represent an approximation to the dynamics and irreducible higher $n$-point correlation functions will build up for times larger than $t_0$ because of the interactions in (\ref{eq:lagrangian}). We restrict ourselves to spatially homogeneous initial conditions such that we can Fourier transform with respect to spatial variables. With 
$\langle \psi_i(x) \rangle = 0$
the one-point function for the $\sigma$-field 
\begin{equation}
\phi(x^0) \equiv \langle \sigma(x) \rangle 
\label{eq:fieldexp}
\end{equation}
at initial time is specified by an initial field amplitude $\phi_0$ as
\begin{equation}
\phi(t_0) = \phi_0 \quad , \quad \partial_{x^0} \phi(x^0)|_{x^0=t_0} = 0 \, .
\label{eq:bosoninitialphi}
\end{equation}
For the initial $\vec{\pi}$ fields we take
\begin{equation}
\langle \vec{\pi}(x) \rangle|_{x^0=t_0} = 0 \quad , \quad \langle \partial_{x^0}\vec{\pi}(x) \rangle|_{x^0=t_0} = 0 \, .
\label{eq:bosoninitialpi}
\end{equation}
It remains to specify two-point corelation functions.
Apart from the above spectral functions, whose initial conditions are fixed by the equal-time relations at initial time, we also have to give the respective commutator expectation values for the fermions and anti-commutators for the bosons. 
These so-called statistical two-point functions are \cite{2PIrefs}
\begin{eqnarray}
F_{\psi,ij}(x,y) &\equiv& \frac{1}{2} \langle [ \psi_i(x), \bar \psi_j(y) ] \rangle \, ,
\label{eq:Fpsi}\\
F_{\sigma}(x,y) &=& \frac{1}{2}\langle \{ \sigma(x), \sigma(y) \} \rangle - \langle \sigma(x) \rangle \langle \sigma(y) \rangle
\label{eq:Fphi}
\end{eqnarray}
and, similarly, the anti-commutator expectation value $F_{\pi}(x,y)$ for each of the $\vec{\pi}$ components. Their spatial Fourier modes at initial time are chosen as
\begin{equation}
F_{\psi,ij}(x^0,y^0,{\bf p})|_{x^0=y^0=t_0}=\frac{m_{\psi}-\gamma^{i} p_i}{\omega_\psi({\bf p})}\left(\frac{1}{2}-n_{\psi}({\bf p})\right) \delta_{ij} \, .
\label{eq:Fpsiinitial}
\end{equation}
with $\omega_\psi({\bf p}) = \sqrt{m_\psi^2+ {\bf p}^2}$. Here the effective fermion mass $m_\psi$ is given by $g \phi_0/2$ at initial time and $n_{\psi}({\bf p})=0$ for vacuum initial conditions. For the initial boson correlators we take
\begin{eqnarray}
F_{\sigma}(x^0,y^0,{\bf p})|_{x^0=y^0=t_0} &=& \frac{1}{\omega({\bf p})} \left(\frac{1}{2}+n({\bf p})\right) \, ,
\nonumber\\
\partial_{x^0}F_{\sigma}(x^0,y^0,{\bf p})|_{x^0=y^0=t_0} & = & 0 \, ,
\nonumber\\
\partial_{x^0}\partial_{y^0}F_{\sigma}(x^0,y^0,{\bf p})|_{x^0=y^0=t_0} & = & \omega({\bf p})
\left(\frac{1}{2}+n({\bf p})\right)
\label{eq:bosoninitial2}
\end{eqnarray}
with $\omega({\bf p}) = \sqrt{m^2 + {\bf p}^2}$ and $n({\bf p})=0$. We choose the same initial conditions for $F_\sigma$ as well as for $F_\pi$. Initial two-point function involving different fields are taken to vanish. The above completely specifies the initial value problem for our model. 

To extract information about particle numbers from numerical simulations it is convenient to define bosonic and fermionic effective particle numbers, both of which are in general not conserved for a interacting system out of equilibrium.\footnote{Of course, there is no unique definition of particle number in an interacting theory if the number is not conserved. It is also nowhere needed in our calculations and only used for interpretation of the results. We use standard definitions that are typically employed to connect to discussions in the context of Boltzmann equations.} For bosons the particle number is associated to the equal-time
statistical propagator $F_\sigma(t,t,{\bf p})$ and quasi-particle
energy $\epsilon_\sigma(t,{\bf p})$ according to \cite{2PIrefs}
\bea
\epsilon_\sigma(t,{\bf p}) &=& \sqrt{\frac{\partial_{t}\partial_{t'}F_\sigma(t,t',{\bf p})\mid_{t=t'}}{F_\sigma(t,t,{\bf p})}} \, ,
\nonumber\\
n{}_\sigma(t,{\bf p})&=&F_\sigma(t,t,{\bf p})\epsilon_\sigma(t,{\bf p})-\frac{1}{2} \, 
\eea
and equivalently for the $\vec{\pi}$ fields.
Plugging the above initial values into these definitions confirms that the vacuum we are starting
from contains no particles according to this definition.

To discuss properties of $F_\psi(t,{\bf p})$ we consider its scalar, pseudoscalar and vector components:
\begin{equation}
F_{S}(t,{\bf p})=\frac{1}{4}{\rm Tr}\left(F_\psi(t,t,{\bf p})\right)\, ,
\end{equation}
\begin{equation}
F_{V}^{i}(t,{\bf p})=\frac{1}{4}{\rm Tr}\left(\gamma^{i}F_\psi(t,t,{\bf p})\right)\, ,
\end{equation}
\begin{equation}
F_{PS}(t,{\bf p})=\frac{1}{4}{\rm Tr}\left(\gamma_{5}F_\psi(t,t,{\bf p})\right)\, ,
\end{equation}
where the trace acts in Dirac space. The flavour indices are omitted here, because we restrict ourselves to initial conditions which are diagonal in flavour space and thus consider 
only flavour-averaged quantities.
Each of these quantities can be used to define an effective particle
number, enabling us to construct $n_{\psi}(t,{\bf p})$ from different
combinations of $F_{S}(t,{\bf p})$, $F_{PS}(t,{\bf p})$ and $F_{V}^{i}(t,{\bf p})$.
Here we follow \cite{BergesGelfandPruschke,Berges:2002wr,PrevFerm} and employ
\begin{equation}\label{eqn:particle_number}
n_{\psi}(t,{\bf p})=\frac{1}{2}-\frac{p_{i}F_{V}^{i}(t,{\bf p})+m_{\psi}(t,{\bf p})F_{S}(t,{\bf p})}{\sqrt{{\bf p}^2+m_{\psi}^{2}(t,{\bf p})}}.
\end{equation}
The time dependence of the effective fermion mass results from the dynamical macroscopic field $\phi(t)$.

\section{Real-time lattice approach}
\label{method}
\subsection{Classical-statistical mapping}

There is a significant class of problems where the dynamics of bosonic quantum fields can be accurately mapped onto a classical-statistical system. This is the case whenever anti-commutator expectation values for bosonic fields are much larger than the corresponding commutators \cite{ClassicalAspects}, which has been studied extensively for scalar field theories~\cite{Khlebnikov:1996mc,ClassicalAspects,MichaTkachevPRL,MichaTkachev,BergesRothkopfSchmidt} as well as pure gauge theories~\cite{BergesSchefflerSexty,FirstGaugeTurbulence,SU3GaugeSim,OverpopulatedGluons,TurbulentSchlichting,NonlinearGlasma}. This classicality condition reads in our case
\begin{eqnarray}
\left( F_{\sigma}(x,y) + \phi(x^0) \phi(y^0) \right)^2 \gg  \rho_\sigma^2(x,y) \,\, , \,\
F_{\pi}^2(x,y) \gg \rho_\pi^2(x,y) \, .
\nonumber\\
\end{eqnarray}
Stated differently, this concerns the large field or large occupancy limit, which is relevant for important phenomena such as nonequilibrium instabilities, particle creation from large coherent fields or wave turbulence which will be relevant for our study. The description breaks down once the typical field occupancies become of order unity. In particular, this is the case in thermal equilibrium. For an introductory review see Ref.~\cite{2PIrefs}. 

In this limit, observables can be obtained as ensemble averages of solutions of classical field equations. 
Suppressing for a moment the $\vec{\pi}$ fields in the notation, one considers canonical field variables at initial time $t_0$, i.e.\ $\sigma_0({\bf x}) = \sigma_{\rm cl}(t_0,{\bf x})$ and $\Pi_0({\bf x})=\partial_t \sigma_{\rm cl}(t,{\bf x})|_{t=t_0}$ for the classical field $\sigma_{\rm cl}(x)$. The values for the canonical field variables at initial time are distributed according to a normalized phase-space density functional $W[\sigma_0,\Pi_0]$, such that an observable $\langle O \rangle$ is given by its phase-space average \cite{ClassicalAspects,ClassicalTruncation}
\begin{equation}
\label{eq:expCS}
\langle O\rangle = \int D \sigma_0\, D \Pi_0~W[\sigma_0,\Pi_0]~O_{\text{cl}}[\sigma_0,\Pi_0]\;. 
\end{equation} 
Here $O_{\text{cl}}[\sigma_0,\Pi_0] = \int D\sigma_0~O[\sigma]~\delta(\sigma-\sigma_{\text{cl}}[\sigma_0,\Pi_0])$,
where $\sigma_{\text{cl}}[\sigma_0,\Pi_0]$ is the solution of the classical field equation with initial conditions $\sigma_0$ and $\Pi_0$. Ensemble averages at initial time are taken to correspond to the respective quantum expectation values for the fields.

Fermions are never largely occupied and are, in this respect, genuinely quantum. However, fermions appear quadratically in the Lagrangian (\ref{eq:lagrangian}) as is also the case for theories like quantum chromodynamics or electrodynamics.\footnote{For all practical purpuses, it can always be achieved that the fermions appear quadratically in the Lagrangian at the expense of introducing extra bosonic field degrees of freedom.} Therefore, their dynamics can be solved without further approximations for given classical bosonic field configuration. 

The procedure is to integrate out the fermions from the path integral to get the classical evolution equation for the bosons. This equations depends then on the fermion currents, represented by fermion two-point correlation functions. The evolution for these fermion correlation functions is obtained from the original Lagrangian, where the fermion fields appear quadratically. This gives a Dirac-like equation for the fermion correlation functions, which is coupled to inhomogeneous classical Bose fields. The description is very suitable for initial value problems and, below, we will find that it accurately describes the quantum dynamics including loop corrections for very non-trivial situations where the latter can be computed using 2PI effective action techniques. 

The above model (\ref{eq:lagrangian}) leads for classical bosonic fields $\sigma_{\text{cl}}(x)$ and $\vec{\pi}_{\text{cl}}(x)$ to the equations of motion
\begin{equation}
\left(\square_{x}+m^{2}\right)\sigma_{\text{cl}}(x)+\frac{\lambda}{4!}\left(\sigma_{\text{cl}}^2 +\vec{\pi}^2_{\text{cl}} \right)\sigma_{\text{cl}}(x)-\frac{g}{2}{\rm Tr}\left( F_\psi(x,x) \right) = 0
\label{eqn:scalar_EoMphi}
\end{equation}
and
\begin{equation}
\left(\square_{x}+m^{2}\right)\vec{\pi}_{\text{cl}}(x)+\frac{\lambda}{4!}\left(\sigma_{\text{cl}}^2 +\vec{\pi}^2_{\text{cl}} \right)\vec{\pi}_{\text{cl}}(x)-\frac{ig}{2} {\rm Tr}\left(F_\psi(x,x)\gamma_{5}\right)=0
\label{eqn:scalar_EoMpi}
\end{equation}
where the trace acts in flavour and Dirac space. These evolution equations depend on the fermion two-point correlator (\ref{eq:Fpsi}). For given classical bosonic fields, the equation of motion for the spinor field $\psi_i(x)$ reads:
\begin{equation}
\left[i\partial_{\mu}\gamma^{\mu}-\frac{g}{2}\left(\sigma_{\text{cl}}(x)+i\gamma_{5}\vec{\tau}\vec{\pi}_{\text{cl}}(x)\right)\right]\psi_i(x)=0.
\label{eq:fermion}
\end{equation}
By multiplying from the right with $\bar \psi_j(y)$ we can continue to construct the evolution equation for the fermion commutator. Since the fermion fields only appear quadratically in the Langrangian, one obtains a Dirac-like equation of motion for the expectation value of the commutator (\ref{eq:Fpsi}):
\begin{equation}\label{Dirac_equations}
\left[i\partial_{x,\mu}\gamma^{\mu}-\frac{g}{2}\left(\sigma_{\text{cl}}(x)+i\gamma_{5}\vec{\tau}\vec{\pi}_{\text{cl}}(x)\right)\right]F_{\psi,ij}(x,y)=0,
\end{equation}

As already mentioned the bosonic fields are treated in the classical-statistical
approximation, where the fields are evolved in time and space according to (\ref{eqn:scalar_EoMphi}) and (\ref{eqn:scalar_EoMpi}) for initial conditions that are sampled to give on average the initial values (\ref{eq:bosoninitialphi}), (\ref{eq:bosoninitialpi}) and (\ref{eq:bosoninitial2}). For each run the coupled system of equations including the one for the fermion two-point function (\ref{Dirac_equations}) with initial condition (\ref{eq:Fpsiinitial}) is solved. In particular, the evolution of all higher bosonic
$n$-point correlation functions can be easily constructed by averaging over
products of $\sigma_{\text{cl}}$ and $\vec{\pi}_{\text{cl}}$ such as $\langle \sigma_{\text{cl}}(x) \sigma_{\text{cl}}(y) \sigma_{\text{cl}}(z) \ldots\rangle$. Usually the number of simulations needed to achieve convergence is reduced by the lattice average for the spatially homogeneous ensemble characterized by the initial conditions (\ref{eq:bosoninitialphi})--(\ref{eq:bosoninitial2}).

\subsection{Fermion dynamics}\label{sec:male-female}

We will describe two methods for a numerical solution of the evolution equation for the fermion correlation function (\ref{Dirac_equations}) and begin in this section with a stochastic method that turns out to be particularly efficient. To actually compute the time evolution of the required statistical
propagator $F_\psi(x,y)$ one has to specify the form of the field
$\psi(x)$ at initial time, where we will suppress the flavor indices in the following. To this end the anticommuting ladder
operators $b_{s}({\bf p})$ for particles and $d_{s}({\bf p})$
for antiparticles will be used in the framework of canonical quantization, where the spinor index $s$ runs from $1$ to $2$. These operators are characterized by their aniti-commutators
\begin{eqnarray}
\left\{ b_{s}({\bf p}),b_{s'}^{\dagger}({\bf q})\right\} =(2\pi)^{3}\delta_{ss'}\delta(\bf{p}-\bf{q})\, ,
\\
\left\{ d_{s}({\bf p}),d_{s'}^{\dagger}({\bf q})\right\} =(2\pi)^{3}\delta_{ss'}\delta(\bf{p}-\bf{q}) \, ,
\end{eqnarray}
while the expectation value of their commutators are parametrized as
\begin{eqnarray}
\left\langle\left[b_{s}({\bf p}),b_{s'}^{\dagger}({\bf q})\right]\right\rangle=(2\pi)^{3}\delta_{ss'}\delta({\bf p}-{\bf q})\left
(1-2n_{+}^{s}({\bf p})\right),
\\
\left\langle\left[d_{s}({\bf p}),d_{s'}^{\dagger}({\bf q})\right]\right\rangle=(2\pi)^{3}\delta_{ss'}\delta({\bf p}-{\bf q})\left
(1-2n_{-}^{s}({\bf p})\right).
\end{eqnarray}
Here $n_{\pm}^{s}({\bf p})$ denote initial occupation numbers of particles and antiparticles of a given spin and spatial momentum. In terms of these operators the initial field $\psi(t_0,{\bf x})$
for our isotropic and homogeneous initial conditions can be written as
\begin{equation}\label{Psi_operator}
\psi(t_0,{\bf x})=\int\frac{d^{3}p}{(2\pi)^{3}}\sum_{s}\left(b_{s}({\bf p})u_{s}({\bf p})e^{-i{\bf p}{\bf x}}+d_{s}^{\dagger}({\bf p})v_{s}(-{\bf p})e^{i{\bf p}{\bf x}}\right).
\end{equation}
The eigenspinors $u_{s}({\bf p})$ and $v_{s}({\bf p})$ represent particle
and antiparticle eigenstates of the Dirac operator. In order
to evaluate $F_\psi(x,y)$ without treating operators explicitly one can either use a mode function expansion \cite{AartsSmit}, which will be discussed below, or utilize the ideas proposed in \cite{BorsanyiHindmarsh} to rewrite the statistical propagator
using a stochastic approach in terms of so-called "male" and "female"
spinor fields $\psi_M(x)$ and $\psi_F(x)$:
\begin{equation}
F_\psi(x,y)=\langle\psi_M(x)\bar{\psi}_F(y)\rangle_{MF}=\langle\psi_F(x)\bar{\psi}_M(y)\rangle_{MF}.
\end{equation}
This procedure of expressing the time evolution
of $F_\psi(x,y)$ in terms of $\psi_M(x)$ and $\psi_F(y)$
is applicable since the equations of motion
for the fermions are linear \cite{BorsanyiHindmarsh}.
The last equation already shows that the roles of "male" and "female"
fields are interchangeable. The notation $\langle \ldots \rangle_{MF}$ emphasizes that in this case the average is performed with respect to an ensemble of male/female pairs. Additionally, one requires that both
of the stochastic spinors obey the Dirac-like equation of motion
in accordance to (\ref{eq:fermion}):
\begin{equation}
\left[i\partial_{\mu}\gamma^{\mu}-\frac{g}{2}\left(\sigma_{\text{cl}}(x)+i\gamma_{5}\vec{\tau}\vec{\pi}_{\text{cl}}(x)\right)\right]\psi_{g}(x)=0.
\label{eq:eompsig}
\end{equation}
The index $g$ (gender) distinguishes here between M (male) and F (female)
fields. To reproduce the initial configuration of
$F_\psi(x,y)$ in terms of $\psi_{M}$ and $\psi_{F}$ the latter
are initialized as
\begin{equation}
\psi_{M,F}(t_0,{\bf x})=\int\frac{d^{3}p}{(2\pi)^{3}}\frac{e^{-i{\bf p}{\bf x}}}{\sqrt{2}}\sum_{s}\left(\xi_{s}({\bf p})u_{s}({\bf p})\pm\eta_{s}({\bf p})v_{s}({\bf p})\right).
\end{equation}
So the only difference between "male" and "female" spinors
is the sign in front of the antiparticle component. Here $\xi_{s}({\bf p})$
and $\eta_{s}({\bf p})$ are random numbers coming from a Gaussian
distribution which are used to simulate the expectation values of
products of the ladder operators $b_{s}({\bf p})$ and $d_{s}({\bf p})$.
Their non-vanishing correlators are linked to initial occupation numbers
\begin{eqnarray}\label{random numbers}
\langle\xi_{s}({\bf p})\xi_{s'}^{\ast}({\bf q})\rangle_{MF}=(2\pi)^{3}\delta_{ss'}\delta({\bf p}-{\bf q})\left(1-2n_{+}^{s}({\bf p})\right),
\\
\langle\eta_{s}({\bf p})\eta_{s'}^{\ast}({\bf q})\rangle_{MF}=(2\pi)^{3}\delta_{ss'}\delta({\bf p}-{\bf q})\left(1-2n_{-}^{s}({\bf p})\right).
\end{eqnarray}
To realize these correlations in a numerical simulation one has to
average over many pairs of "male" and "female" fields, but
of course each of them has to be evolved in time separately. If the
number of pairs is sufficiently large the result will converge to
the physical correlator. Later we will discuss
how many male/female pairs are actually required. 

Bilinears such as $F_\psi(x,y)$ are
computed by combining spinors of both genders and the statistical propagator is evolved in time by solving the Dirac equations for $\psi_M$ and $\psi_F$. It is illustrative to compute the initial 
$F_\psi(x^0,y^0,{\bf p})|_{x^0=y^0=t_0}$ in terms
of operator-valued field $\psi$ at initial time and the same quantity
from $\psi_{M}$ and $\psi_{F}$. Both calculations yield for symmetric
occupation numbers $n_{\pm}({\bf p})=n_{\pm}(-{\bf p})$:
\begin{eqnarray}
F_\psi(x^0,y^0,{\bf p})|_{x^0=y^0=t_0}&\! = \!& \frac{1}{2} \sum_{s}\Big[\left(1-2n_{+}^{s}({\bf p})\right)u_{s}({\bf p})\bar{u}_{s}({\bf p})
\nonumber\\
&&-\left(1-2n_{-}^{s}({\bf p})\right)v_{s}({\bf p})\bar{v}_{s}({\bf p})\Big] \, .
\end{eqnarray}

In the following, we describe the standard mode-function approach for comparison. The starting point of the mode-function expansion is again a Fourier expansion of the fermionic field operator at $t=t_0$ as in (\ref{Psi_operator}), which can be generalized to $t > t_0 $ by introducing time-dependent mode-functions $\Phi^u_s (t,{\bf x},{\bf k})$ and $\Phi^v_s (t,{\bf x},{\bf k})$: 
\begin{equation} \label{modefuncexp}
\psi(t,{\bf x}) = \int \frac{d^{3}p}{(2\pi)^{3}} \sum_{s} \left( b_s ({\bf p}) \Phi^u_s (t,{\bf x},{\bf p}) 
         + d_s^{\dagger} ({\bf p}) \Phi^v_s (t,{\bf x},-{\bf p}) \right).
\end{equation}
Here $b_s({\bf p})$ and $d_s({\bf p})$ are again the 
annihilation operators at initial time and we have
\begin{equation}
\Phi^u_s (t_0,{\bf x},{\bf p}) = u_{s}({\bf p})e^{-i{\bf p}{\bf x}} \,\, , \,\, 
\Phi^v_s (t_0,{\bf x},{\bf p}) = v_{s}({\bf p})e^{-i{\bf p}{\bf x}} \, .
\end{equation}
Substituting (\ref{modefuncexp}) into the equation of motion for $\psi(x)$, one 
observes that every mode function has to satisfy 
\begin{equation}
\left[i\partial_{\mu}\gamma^{\mu}-\frac{g}{2}\left(\sigma_{\text{cl}}(x)+i\gamma_{5}\vec{\tau}
\vec{\pi}_{\text{cl}}(x)\right)\right] \Phi^{u/v}_s(x,{\bf p}) = 0 \, .
\end{equation}
After the time evolution of $ \Phi^{u/v}_s(x,{\bf p})$ has been
calculated, observables can be constructed using the expansion
(\ref{modefuncexp}). For the evaluation of the expectation values one 
has to use properties of the initial state such as 
\bea
 \left\langle b_i^{\dagger} ({\bf p}) b_i({\bf p})
 \right\rangle =n^u_i (t=0,{\bf p}) , \ \ 
 \langle b_s({\bf p}) \rangle = 0 , \ \ 
\langle b^{\dagger}_s({\bf p}) \rangle = 0 , \ \ 
\eea
and similarly for $d_i$.

The statistical two-point function (\ref{eq:Fpsi}) reads
in terms of mode functions 
\bea
F_\psi(x,y)& = & \frac{1}{2}\int {d^3 p \over 2 \pi} \sum_s \Big( \left\langle b_i b_i^{\dagger} - b_i^{\dagger} b_i \right\rangle 
\Phi_s^u(x,{\bf p}) \bar \Phi_s^u (y,{\bf p}) 
\nonumber\\
&& + \left\langle d_i^{\dagger} d_i - d_i d_i^{\dagger} \right\rangle 
\Phi_s^v(x,{\bf p}) \bar \Phi_s^v (y,{\bf p}) \Big) \nonumber \\
&= & 
\int {d^3 p \over 2 \pi} \sum_s
\left( \frac{1}{2}- n^u _{in,s}({\bf p}) \right) \Phi_s^u(x,{\bf p}) \bar \Phi_s^u (y,{\bf p}) \nonumber \\
 &&   + \int {d^3 p \over 2 \pi} \sum_s
\left(n^v_{in,s}({\bf p}) -\frac{1}{2} \right)  \Phi_s^v(x,{\bf p}) \bar 
\Phi_s^v (y,{\bf p}). \nonumber \\ 
\label{fwithmodefunc}
\eea
Here the subscript "in" stresses the fact that the particle numbers appearing here are evaluated at the initial time. Similarly, we can also calculate the fermion contribution to the energy density of the system $\langle H_D(x) \rangle$ with the Dirac Hamiltonian $ H_D = - i \gamma^0 \gamma^i \partial_i + \gamma^0 m $ and any other observable $O(x)$ of interest by a summation over the 
mode functions: $\sum_i F_i \bar \Phi_i O(x) \Phi_i$, 
where the index $i$ 
represents the momentum, spin and charge of the mode, and the summation over 
the index $i$ represents the momentum integral and the sum over 
spin and charge modes, and $F_i$ depends on $ n^u_{in,i}$ and $ n^v_{in,i} $, 
similarly to (\ref{fwithmodefunc}).

 The advantage of this method is that it is exact
without further approximations and involves no ensemble average as the
male/female approach. The great disadvantage, which so far had limited
its applicability to lower-dimensional systems, is the requirement to
simulate a mode function for every possible combination of space and
momentum. If implemented on a lattice, as is described in the
following, the mode-function method leads to prohibitively high
computational costs on bigger lattices.

We finally note that one can also build another low cost noisy ensemble of linear combinations of mode functions using 
\bea
\varphi_j = \sum_i \xi_{j,i} \Phi_i, 
\eea
where the $j=1..N_{ens}$ index labels the mode functions in the ensemble 
and the index $i$ goes over all the modes in the mode function 
expansion (i.e, $i$ represents their momenta, spin and charge),
and  $ \xi_{j,i}$ are random numbers yet to be determined.
Since the equation of motion for the mode functions is linear, any linear 
combination will be a solution as well. 

If we choose the random numbers such that 
\bea 
 \langle \xi^*_{j,i} \xi_{l,k} \rangle = F_i \delta_{ik} \delta_{jl}
\eea
then the ensemble average 
\bea {1\over N_{ens}} \sum_{j=1}^{N_{ens}}
\langle \bar{\varphi}_j O(x) \varphi_j \rangle 
= \sum_i F_i \bar{\Phi}_i O(x) \Phi_i
\eea
gives the result of the mode function expansion. 
(In practice one needs two ensembles to be able to calculate 
any observable, corresponding to 
the two possible orderings of the ladder operators in the observable.)

This formulation of the noisy ensemble performs similarly to the 
male/female formulation, such that the expectation value can be estimated 
using much less ensemble members $ \varphi_j$ than a corresponging mode
function simulation would use (in more than one spatial dimension).

\subsection{Lattice formulation}\label{sec:lattice}

Numerical simulations based on the methods described
in the last subsections are carried out using a leap-frog algorithm
on a $3+1$ dimensional space-time lattice with a real time coordinate.
The simulated spatial volume is $V=(Na_{s})^{3}$
with spatial lattice spacing $a_{s}$ and number of lattice points $N$
in each direction, where periodic spatial boundary conditions are used. 
The momentum resolution is determined by $N$ and $a_{s}$,
with the highest possible lattice momentum proportional to $1/a_{s}$
and the lowest one to $1/(N a_s)$. The time direction
is discretized using a lattice spacing $a_{t}$ and length $t_{\rm max}=N_{t}a_{t}$.\footnote{There is no (anti-)periodicity in real time employed, as is typically the case for Euclidean lattice fomulations.} 

First- and second-order derivatives on the lattice are
discretized using symmetric finite difference approximations 
\bea
f'(x)=\frac{f(x+a_{i})-f(x-a_{i})}{2a_{i}} \, ,
\nonumber\\
f''(x)=\frac{f(x+a_{i})+f(x-a_{i})-2f(x)}{a_{i}^{2}}.
\eea
Here $a_{i}$ with $i=s,t$ denotes either the spatial or temporal lattice spacing. For
the calculation of second-order bosonic spatial derivatives an alternative
higher-order discretization has been applied, which is slightly more
accurate for larger lattice spacings:
\bea
f''(x) &=& \frac{16f(x+a_{i})+16f(x-a_{i})-f(x+2a_{i})}{12a_{i}^{2}}
\nonumber\\
&&-\frac{f(x-2a_{i})+30f(x)}{12a_{i}^{2}} \, .
\label{eq:higherdiscret}
\eea

To remain consistent
with the lattice version of derivatives it is also necessary to redefine
lattice momenta, which is achieved by applying discrete derivatives
to plane-wave solutions, such as $\partial_{x}e^{ipx}=\left[(e^{ipa_{s}}-e^{-ipa_{s}})/(2a_{s})\right]e^{ipx}$.
The corresponding lattice momentum definitions are\footnote{On the lattice momentum integrals are replaced by sums: $\int\frac{d^3p}{\left(2\pi\right)^3}\longrightarrow\frac{1}{N^3a_s^3}\sum_{\bf{p}}$.} 
\begin{equation}
\bar p_{i}=\frac{\sin(p_{i}a_{s})}{a_{s}}  \quad , \qquad i=1,2,3
\end{equation}
for a first order spatial derivative as appearing in a Dirac equation, or 
\begin{equation}
{\bf p}_{\rm lat}^{2}=\frac{1}{a_{s}^{2}}\sum_{i=1}^{3}4\sin^{2}\left(\frac{p_{i}a_{s}}{2}\right)
\end{equation}
for second-order spatial derivatives as ones appearing in a Klein-Gordon equation with $p_{i}=2\pi n_{i}/(Na_{s})$ and $n_{i}=0,...,N-1$. For the bosonic sector of our model we use the discretization (\ref{eq:higherdiscret}), which leads to
\begin{equation}
{\bf p}_{\rm lat}^2 =\frac{1}{a_{s}^{2}}\sum_{i=1}^{3}\left[2.5-\frac{8}{3}\cos\left(p_{i}a_{s}\right)+\frac{1}{6}\cos\left(2p_{i}a_{s}\right)\right]\, .
\end{equation}

For fermionic degrees of freedom such a
straightforward discretization is
known to cause so-called fermion doublers \cite{doublers}. A way to address this
problem, similar to the one commonly employed in Euclidean lattice gauge theory, is to introduce a spatial Wilson term $W$ into the equation of motion (\ref{eq:eompsig}):
\begin{equation}
\left[i\partial_{\mu}\gamma^{\mu}+W-\frac{g}{2}\left(\sigma(x)+i\gamma_{5}\vec{\tau}\vec{\pi}(x)\right)\right] \psi_g(x)=0 \, .
\end{equation}
A standard choice would be
\begin{equation}
W \psi_g(x) =\frac{ra_{s}}{2}\bigtriangleup_x\psi_g(x)\, ,
\end{equation}
where we set $r = 1$ from now on and use the Laplacian
\begin{equation}
\bigtriangleup_x\psi_g(x) = \sum_{i=1}^{3}\frac{\psi_g(x+a_{i})+\psi_g(x-a_{i})-2\psi_g(x)}{a_{s}^{2}}
\end{equation}
which leads to a momentum-dependent contribution to the fermionic dispersion relation: 
\begin{equation}
\omega({\bf p})=\sqrt{m^{2}_{\psi}+\bar p_i \bar p^i+a_{s}m_{\psi} {\bf p}_{\rm lat}^{2}+\frac{a^{2}_{s}}{4}{\bf p}_{\rm lat}^{4}} \, .
\end{equation}
This ensures that only low-momentum excitations show a low-energy dispersion relation. One observes that the additional contributions from the Wilson term vanish in the continuum limit $a_s\to0$. However, we find that a faster approach to the continuum limit is achieved by replacing $W \rightarrow W_{PS}$ defined as 
\begin{equation}
W_{PS} \psi_g(x) =i\gamma_5\frac{ra_{s}}{2}\bigtriangleup_x\psi_g(x).
\end{equation}
Here the subscript $PS$ means pseudoscalar in contrast to the standard scalar Wilson term. The pseudoscalar Wilson term also leads to a (modified) momentum dependent contribution
to the fermionic dispersion relation 
\begin{equation}\label{dispersion_PS}
\omega({\bf p})=\sqrt{m^{2}_{\psi}+\bar p_i \bar p^i+\frac{a^{2}_{s}}{4}{\bf p}_{\rm lat}^{4}}.
\end{equation}
This illustrates that $W_{PS}$ eliminates the $O(a_s)$ contribution to the dispersion relation which will be particularly advantageous for our out-of-equilibrium setup where the effective fermion mass will be time-dependent.\footnote{A related construction used in lattice gauge theory is known as twisted mass fermions.}
We do not include a temporal Wilson term as this would turn a Dirac equation into a second order differential equation in time. 
The temporal doublers are avoided provided that we initialize only the physical mode and choose the temporal lattice spacing to be much smaller than the spatial lattice spacing $a_t\ll a_s$ \cite{AartsSmit,BorsanyiHindmarsh,SaffinTranberg1,SaffinTranberg2,MouSaffinTranberg}.

At the beginning of each simulation initial conditions
have to be specified. For the bosons one has to specify the initial classical fields 
and derivatives according to (\ref{eq:bosoninitialphi}), (\ref{eq:bosoninitialpi}) and (\ref{eq:bosoninitial2}) to obtain quantum-like vacuum initial conditions. 
They are initialized
in momentum space from a Gaussian probability distribution centered
around zero with standard deviation of $\left(2\sqrt{m^{2}+({\bf p})^{2}}\right)^{-1/2}$
for the fields and $\left(\sqrt{m^{2}+{\bf p}^{2}}/2\right)^{1/2}$
for the time derivatives of the fields. In order to realize parametric resonance in quantum field theory the average initial field is homogeneous with amplitude 
\begin{equation}
\phi(t=0)=\sigma_0\sqrt{\frac{6N}{\lambda}} \, ,
\label{eq:sigma0} 
\end{equation}
where the parameter $\sigma_0$ sets the overall scale for our simulations.

Another important property in momentum
space is $\sigma(-{\bf p})=\sigma^*({\bf p})$ and similarly for $\vec{\pi}$, which is
required to get real-valued fields in position
space. This is achieved by multiplying the real field amplitudes with a random phase
factor $e^{i\alpha(\bf{p})}$ with $\alpha({\bf p})=-\alpha(-{\bf p})$
and $\alpha= 0$ for $p_i = 0$ and $p_i = \pi/a_{s}$.
The same procedure using another random phase factor $e^{i\beta(\bf{p})}$
is applied to get real-valued field derivatives w.r.t. time in position
space. Due to the fact that in our approach the scalar fields are classical-statistical we try to minimize possible effects of UV divergent contributions by initializing the quantum-like vacuum correlators only up to a finite momentum $\Lambda<|{\bf p}|_{\rm lat}^{\rm max}$, with $\Lambda$ being higher than all of the momentum modes which become relevant during the simulated time.

For the low-cost fermion approach, initial values are given in terms of
$\psi_M(x)$ and $\psi_F(x)$ at $t=t_0=0$. They are directly
linked to complex random numbers, which have to fulfill correlator
relations of the fermionic ladder operators (\ref{random numbers}). To start with vacuum initial conditions we set all particle numbers to zero. These correlator relations are implemented numerically through complex
valued $\xi_{s}({\bf p})=A_{s}({\bf p})e^{i\varphi_{s}({\bf p})}$
and $\eta_{s}({\bf p})=B_{s}({\bf p})e^{i\theta_{s}({\bf p})}$ with
real amplitudes $A_{s}({\bf p})$ and $B_{s}({\bf p})$ coming from
a Gaussian distribution and random phases $\varphi_{s}({\bf p})$
and $\theta_{s}({\bf p})$ to ensure that all mixed correlators vanish.
Having chosen a symmetric finite difference approximation for the
first time derivative, we have to specify $\psi_M(x)$ and
$\psi_F(x)$ not only at $t=0$ but also at $t=-a_{t}$, which
is done by an evolution of the free fields according to 
\begin{equation}
\psi_{M,F}(t=0,{\bf p})=e^{-i\gamma_0\omega({\bf p})a_t}\psi_{M,F}(t=-a_{t},{\bf p}).
\end{equation}

The initial statistical propagator $F_\psi(x,y)|_{x^0=y^0=0}$, which solves the free Dirac equation at $t=0$, reads on the lattice in the presence of the employed Wilson term 
\begin{equation}
F_\psi(x^0=0,y^0=0,{\bf p})=\frac{m_{\psi}-\gamma^{i}\bar p_i-i\gamma_5\frac{a_s}{2}{\bf p}_{\rm lat}^2}{2\omega({\bf p})}\left(1-2n_{\psi}({\bf p})\right).
\end{equation}
Likewise, the fermion occupation number (\ref{eqn:particle_number}) is given by
\begin{equation}
n_{\psi}(t,{\bf p})=\frac{1}{2}-\frac{\bar p_{i}F_{V}^{i}(t,{\bf p})+m_{\psi}(t,{\bf p})F_{S}(t,{\bf p})+i\frac{a_s}{2}{\bf p}^2_{\rm lat}F_{PS}(t,{\bf p})}{\sqrt{\bar p_i \bar p^i+m_{\psi}^{2}(t,{\bf p})+\frac{a^{2}_{s}}{4}{\bf p}_{\rm lat}^{4}}}.
\label{eq:npsilat}
\end{equation}

\subsection{Renormalization}\label{renorm}

\begin{figure}[!t]
\begin{center}
\epsfig{file=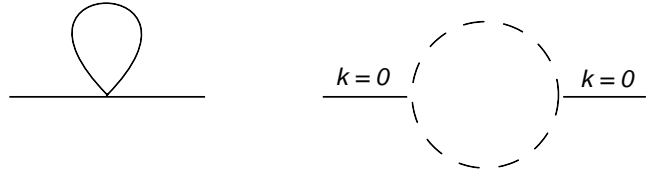 , width = \mywidth }
\caption{Left: Scalar tadpole. Right: Fermion loop with vanishing external momentum.}
\label{fig:renorm_diagrams}
\end{center}
\end{figure}

To obtain physically relevant information from our numerical simulations, we have to ensure that the results are insensitive to changes of the finite lattice cut-off $\sim 1/a_s$. In practice the variation of the cutoff-scale in simulations barely exceeds one order of magnitude. Here we consider the leading divergences perturbatively, which we explicitly verified to lead to cutoff insensitive numerical results for a variation of the spatial lattice spacing in the range $a_s \sigma_0 = 0.1$ -- $1$. For our model this concerns the quadratically running scalar mass terms, where the relevant contributions are coming from the one-loop scalar self-energy corrections displayed in Fig.~\ref{fig:renorm_diagrams}.
Because of our choice of initial conditions with a non-zero $\sigma$-field amplitude (\ref{eq:bosoninitialphi}), the dressings of $\sigma$ and $\vec{\pi}$ masses through vacuum fluctuations are in general different. For given renormalized mass squared $m^2$, we compute the mass parameters $m_{0,\sigma/{\pi}}^2$ self-consistently from
\begin{equation}\label{eqn:mass_tadpole}
m_{0,\sigma/{\pi}}^2+\Sigma_{0,\sigma/{\pi}}(m_{0,\sigma}^2,m_{0,{\pi}}^2)=m^2
\end{equation}
using the analytical form of the self-energies displayed in Fig.~\ref{fig:renorm_diagrams}:
\begin{multline}
\Sigma_{0,\sigma}=\frac{\lambda}{48}\frac{1}{N^3a_s^3}\sum_{\bf{p}}\left(\frac{3}{\sqrt{m_{0,\sigma}^2+{\bf p}_{\rm lat}^2}}+\frac{3}{\sqrt{m_{0,{\pi}}^2+{\bf p}_{\rm lat}^2}}\right)\\-g^2\frac{1}{N^3a_s^3}\sum_{\bf{p}}\frac{\bar p_i \bar p^i+\frac{a^{2}_{s}}{4}{\bf p}_{\rm lat}^{4}}{\left(\bar p_i \bar p^i+m_{\psi}^2+\frac{a^{2}_{s}}{4}{\bf p}_{\rm lat}^{4}\right)^{3/2}},
\end{multline}
\begin{multline}
\Sigma_{0,{\pi}}=\frac{\lambda}{48}\frac{1}{N^3a_s^3}\sum_{\bf{p}}\left(\frac{1}{\sqrt{m_{0,\sigma}^2+{\bf p}_{\rm lat}^2}}+\frac{5}{\sqrt{m_{0,{\pi}}^2+{\bf p}_{\rm lat}^2}}\right)\\-g^2\frac{1}{N^3a_s^3}\sum_{\bf{p}}\frac{\bar p_i \bar p^i+m_{\psi}^2}{\left(\bar p_i \bar p^i+m_{\psi}^2+\frac{a^{2}_{s}}{4}{\bf p}_{\rm lat}^{4}\right)^{3/2}}.
\end{multline}
At the beginning of each simulation of the real-time dynamics, these equations are first solved by iteration starting with $m_{0,\sigma/{\pi}}^2=m^2$ until convergence is achieved. Then $m^2$ is replaced by $m_{0,\sigma}^2$ in equation (\ref{eqn:scalar_EoMphi}) and by $m_{0,\pi}^2$ in (\ref{eqn:scalar_EoMpi}).\footnote{The Lagrangian (\ref{eq:lagrangian}) describes massless fermions. A non-zero mass parameter for the fermions in the Lagrangian would lead to a divergent linear contribution to the bosonic potential requiring an additional renormalization.}

\subsection{Comparison of male/female and mode function approach}\label{mode functions}

The male/female method described above has to converge to the results
of the mode function expansion in the limit where the number of male/female pairs is sufficiently large.
In practice, the convergence depends on parameters 
such as the dimension $d$ of space, the number of lattice points $N$ 
or value of couplings. In general, simulations employing the mode function expansion are limited 
to relatively small lattices because the number of mode functions increases like $N^{d}$ for every lattice point, therefore the total cost of the simulation increases as $N^{2d}$. In contrast, the cost for male/female fermion simulations is expected to scale as $N^d$ times the number of male/female pairs.
As a consequence, in one spatial dimension the male/female method has no particular advantage over the mode functions approach, as the needed number of pairs is not significantly lower than the number of mode functions per lattice point. The situation is different in two or three dimensions, where one typically observes reasonable convergence for a much lower number of pairs as compared to the requirements of the full mode function expansion.

\begin{figure}[t!]
\begin{center}
\epsfig{file=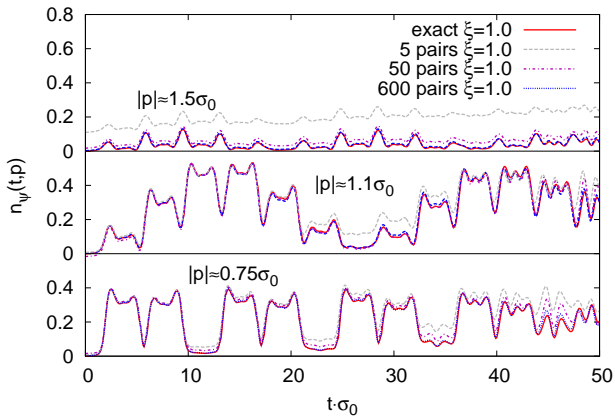, width=\mywidth, angle=0}
\caption{Occupation number of three momentum modes as a function of time and the convergence with increasing number of male/female pairs.}
\label{fig:modef2}
\end{center}
\end{figure}

To give an explicit example, we compare the time-evolution on a small $16^3$ lattice using the full mode function
expansion and the male/female method for varying number of pairs. In the remainder of this work the latter method will then be used on larger lattices to compute results for the analysis of the underlying physics.
In Fig.~\ref{fig:modef2} the fermion occupation number (\ref{eq:npsilat}) is shown as a function of time for different values of the spatial momentum ${\bf p}$. The underlying physical processes will be discussed in detail below. For the plot the number of male/female pairs is varied from $5$ to $600$. We clearly observe that both approaches agree to very good accuracy for a sufficiently large number of pairs. The convergence is typically faster for low momentum modes, in agreement with the expectation that self-averaging is more efficient for low momenta since the involved characteristic volume is larger.  

\begin{figure}[t!]
\begin{center}
\scalebox{0.7}{\epsfig{file=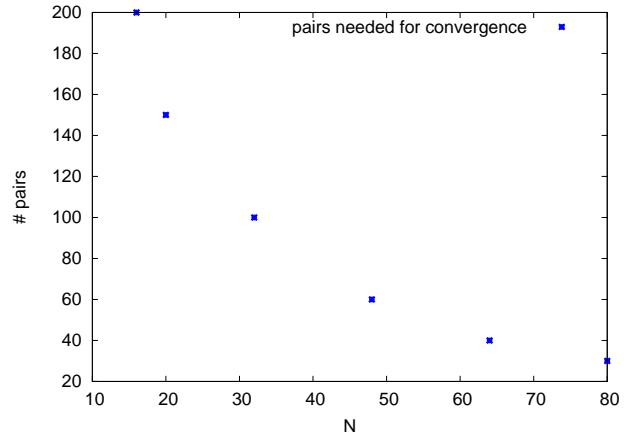, width=\mywidth, angle=270}}
\caption{Number of male/female pairs required to achieve convergence as a function of number of lattice sites in each direction.}
\label{fig:number_pairs}
\end{center}
\end{figure}

For the employed small lattice size in this example we required a relatively high number of pairs. This is expected to change for larger lattices due to enhanced self-averaging. Since a direct comparison with mode function results is impractical for larger lattices, we further investigated the convergence of results as the number of male/female pairs is increased for given lattice sizes. Fig.~\ref{fig:number_pairs} shows the number of male/female pairs that are required to achieve convergence of momentum dependent particle spectra with all other parameters, like simulation time, lattice spacing etc., fixed.  
One observes that for increasing lattice sizes the number of pairs can be reduced. Convergence of momentum independent observables like energy density can be usually be reached with even lower statistics, because of the full employment of self-averaging for these observables. In general, we find that a larger ultraviolet cutoff or stronger coupling worsen the convergence making, in particular, the study of strongly correlated fermions computationally more expensive than weakly correlated systems. 

\section{Validation against quantum field theory}
\label{2PI}

Above we described lattice methods that give a fully non-perturbative description of the dynamics in their range of validity for large bosonic field amplitudes or occupancies. In principle, there are no further approximations in the fermion sector and the lattice description includes the physics of fermion loop corrections to infinite order. However, using Wilson fermions there are additional procedures to suppress fermion doublers on the lattice. It is, therefore, illustrative to validate the lattice description against (continuum) quantum field theory at least in the weak-coupling limit, where this is possible since suitable approximations exist for the latter. The quantum decsription we employ is based on a resummed large-$N$ expansion to next-to-leading order (NLO) for the bosonic sector and a resummed loop expansion for the fermionic sector of our model~\cite{Berges:2001fi,Berges:2002wr}. 

The resummation is efficiently formulated in terms of the two-particle irreducible (2PI) effective action in Minkowski space-time \cite{2PIrefs}
\begin{eqnarray}
\Gamma[\phi,G,G_\psi] &=& S[\phi] + \frac{i}{2}\text{Tr}\ln\left(G^{-1}\right) + \frac{i}{2}\text{Tr}\left(G_0^{-1}(\phi) G\right) 
\\
&-& i\text{Tr}\ln\left(G^{-1}_\psi\right) - i\text{Tr}\left(G_{0,\psi}^{-1}G_\psi\right) + \Gamma_{2}[\phi,G,G_\psi]\nonumber \, ,
\label{eq:Gamma2PI}
\end{eqnarray}
which includes all quantum corrections if the two-particle irreducible part $\Gamma_2$ is known.
Here $\phi(x)$ denotes the field expectation value (\ref{eq:fieldexp}) while $G = {\rm diag}\{ G_\sigma,\vec{G}_\pi\}$ and $G_\psi$ denote the full boson and fermion propagators, which are diagonal in field index space. The traces involve the sum over field indices as well as space-time integrals. The fields live on a closed time path or Schwinger-Keldysh contour ${\mathcal C}$, which runs back and forth along the real-time axis starting at given initial time \cite{2PIrefs}. The classical action part reads 
\begin{equation}
S [\phi] = \int_C dt \int d^3x\, \left( \frac{1}{2} \partial_{\mu}\phi\, \partial^{\mu} \phi - \frac{1}{2}\, m^{2} \phi^2 - \frac{\lambda}{4! N_s}\, \phi^4 \right) \, ,
\end{equation}
while the classical propagators are
\begin{eqnarray}
i G_{0,\sigma}^{-1}(x,y;\phi) &=& \frac{\delta^2 S}{\delta \phi(x) \delta \phi(y)}
\nonumber\\
&=& - \left( \square + m^2 + \frac{\lambda}{2 N_s} \phi^2(x) \right) \delta(x-y) \, ,
\nonumber\\
i G_{0,\pi}^{-1}(x,y;\phi) &=& - \left( \square + m^2 + \frac{\lambda}{6 N_s} \phi^2(x) \right) \delta(x-y) \, ,
\nonumber\\
i G_{0,\psi}^{-1}(\phi) &=& \left( i \partial_\mu \gamma^\mu - \frac{g}{2}\, \phi(x) \right) \delta(x-y) \, .
\end{eqnarray}
The real-time quantum evolution equations for $\phi$, $G$ and $G_\psi$ are obtained from (\ref{eq:Gamma2PI}) by
\begin{equation}
\frac{\delta\Gamma[\phi,G,G_\psi]}{\delta \phi(x)} \,=\, 0 \, , \,
\frac{\delta\Gamma[\phi,G,G_\psi]}{\delta G(x,y)} \,=\, 0 \, , \,
\frac{\delta\Gamma[\phi,G,G_\psi]}{\delta G_\psi(x,y)} \,=\, 0 \, ,
\end{equation}
which are solved numerically by discretizing the equations on a sphere in spatial momentum space using standard techniques. In particular, this description requires no Wilson term to remove fermion doublers inherent in the above lattice approach. 

\begin{figure}[t!]
 \begin{center}
 \includegraphics[scale=0.33]{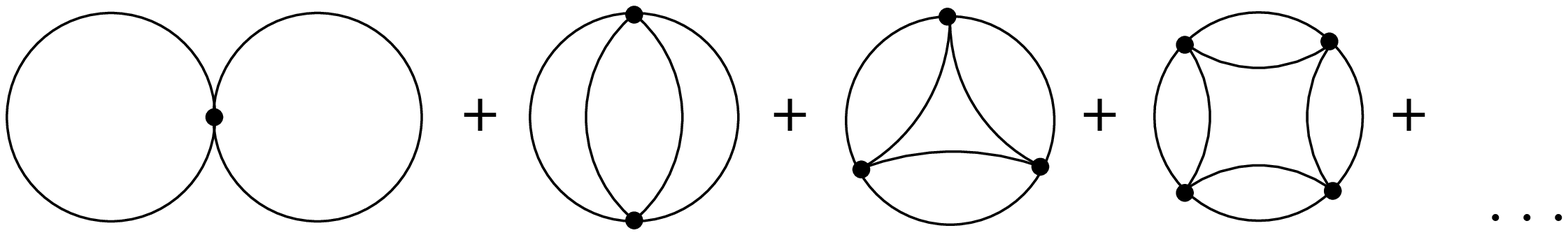}\\
\medskip
\includegraphics[scale=0.33]{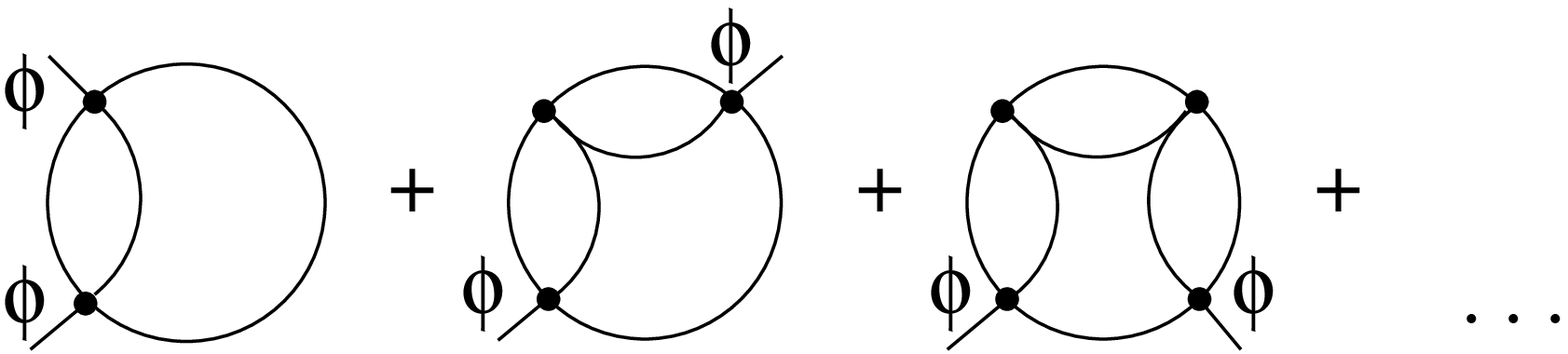}\\
\end{center}
\caption{Expansion of $\Gamma_2$ to NLO in $1/N$ for the bosonic sector. Solid lines are full boson propagators while external legs correspond to insertions of the macroscopic field $\phi$. The dots indicate that we sum up an infinite series of diagrams, with every next diagram having one additional loop in the bubble ring.\label{fig:bosons}} 
\end{figure}

\begin{figure}[t!]
 \begin{center}
 \includegraphics[scale=0.4]{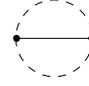}
\end{center}
\caption{Expansion of $\Gamma_2$ to two loops in the fermion sector. A dashed line represents a full fermion propagator.\label{fig:fermions}} 
\end{figure}

Our approximation for $\Gamma_{2}$ is depicted graphically in Figs.~\ref{fig:bosons} and \ref{fig:fermions}. The employed $1/N$ expansion to NLO in the bosonic sector corresponds to summing an infinite series of diagrams~\cite{Berges:2001fi}, while the fermion corrections are taken into account at two-loop order. We will call this approximation 'NLO 2PI' in the following. 
For the comparison, we employ weak couplings $\lambda \ll 1$ and $g \ll 1$. The $1/N$ expansion can describe even non-perturbatively large occupancies of order $1/\lambda$, which will be relevant for the dynamics in the bosonic sector, whereas the occupancies in the fermion sector are strictly limited by the Fermi statistics. Of course, the loop expansion of $\Gamma_2$ in the fermion sector is not expected to be valid for strong couplings. As a consequence, the quantum results can be used to validate the lattice approach for weak coupling only. 

The comparison of quantum and classical-statistical lattice results has been performed in great detail for purely bosonic theories in the past~\cite{ClassicalAspects,ClassicalTruncation}. Here we concentrate on the fermion sector entending our earlier results \cite{BergesGelfandPruschke}. We decompose the propagators into their respective statistical and spectral components as \cite{2PIrefs}
\begin{eqnarray}
G_{\sigma/\pi}(x,y) &=& F_{\sigma/\pi}(x,y)-\frac{i}{2}\rho_{\sigma/\pi}(x,y)\, \mbox{sgn}(x^0-y^0) \, ,
\\
G_\psi(x,y) &=& F_\psi(x,y)-\frac{i}{2}\rho_{\psi}(x,y)\, \mbox{sgn}(x^0-y^0) \, .
\end{eqnarray}
The statistical two-point functions $F_{\sigma/\pi}(x,y)$ and $F_\psi(x,y)$ as well as the corresponding spectral functions
are the ones defined in section (\ref{model}). In particular, we employ the same definitions for extracting the time evolution of particle numbers. 

\begin{figure*}[t!]
  \begin{center}
    \begin{tabular}{cc}
      \resizebox{85mm}{!}{\includegraphics[width=\mywidth, angle=270]{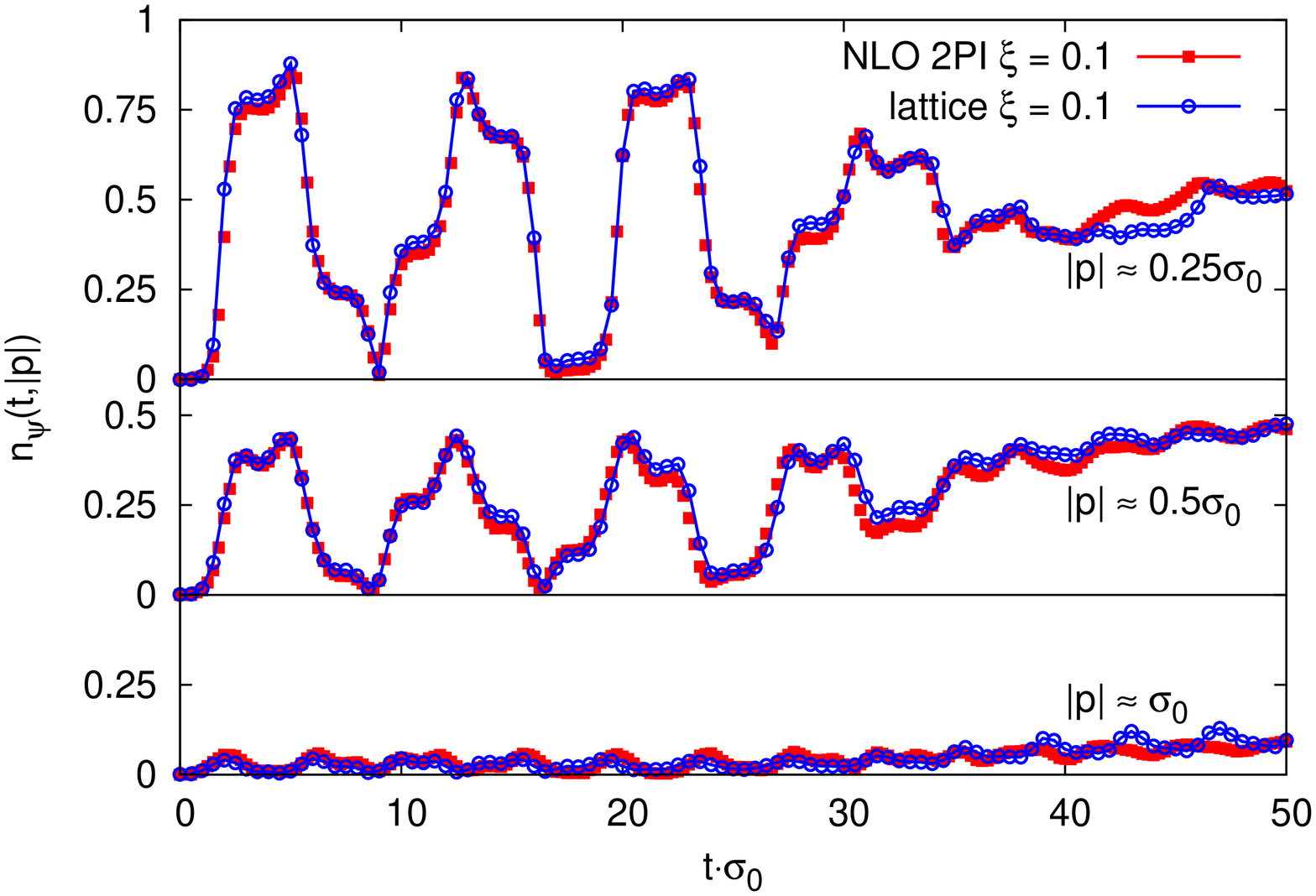}} &
      \resizebox{85mm}{!}{\includegraphics[width=\mywidth, angle=270]{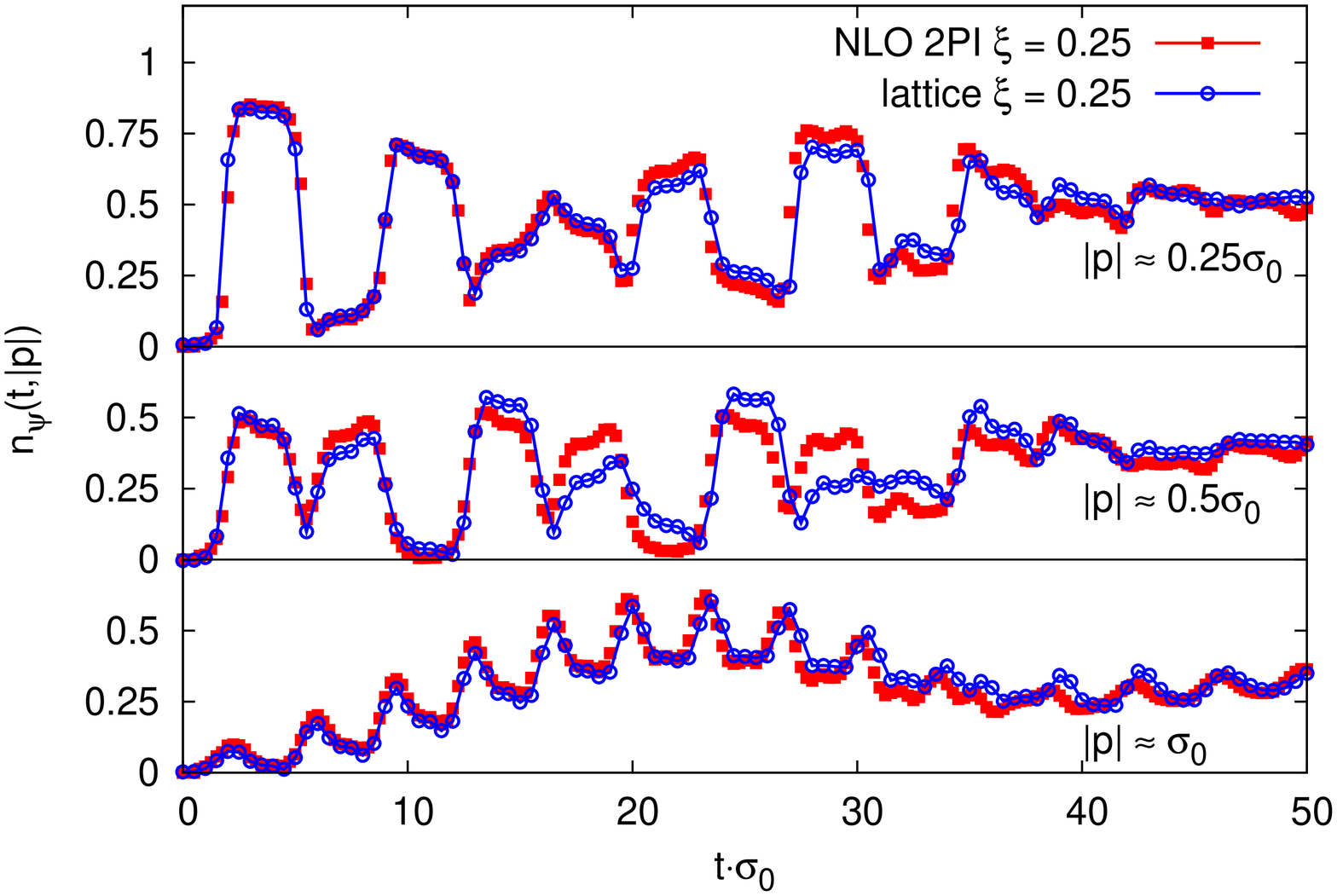}} \\
      \resizebox{85mm}{!}{\includegraphics[width=\mywidth, angle=270]{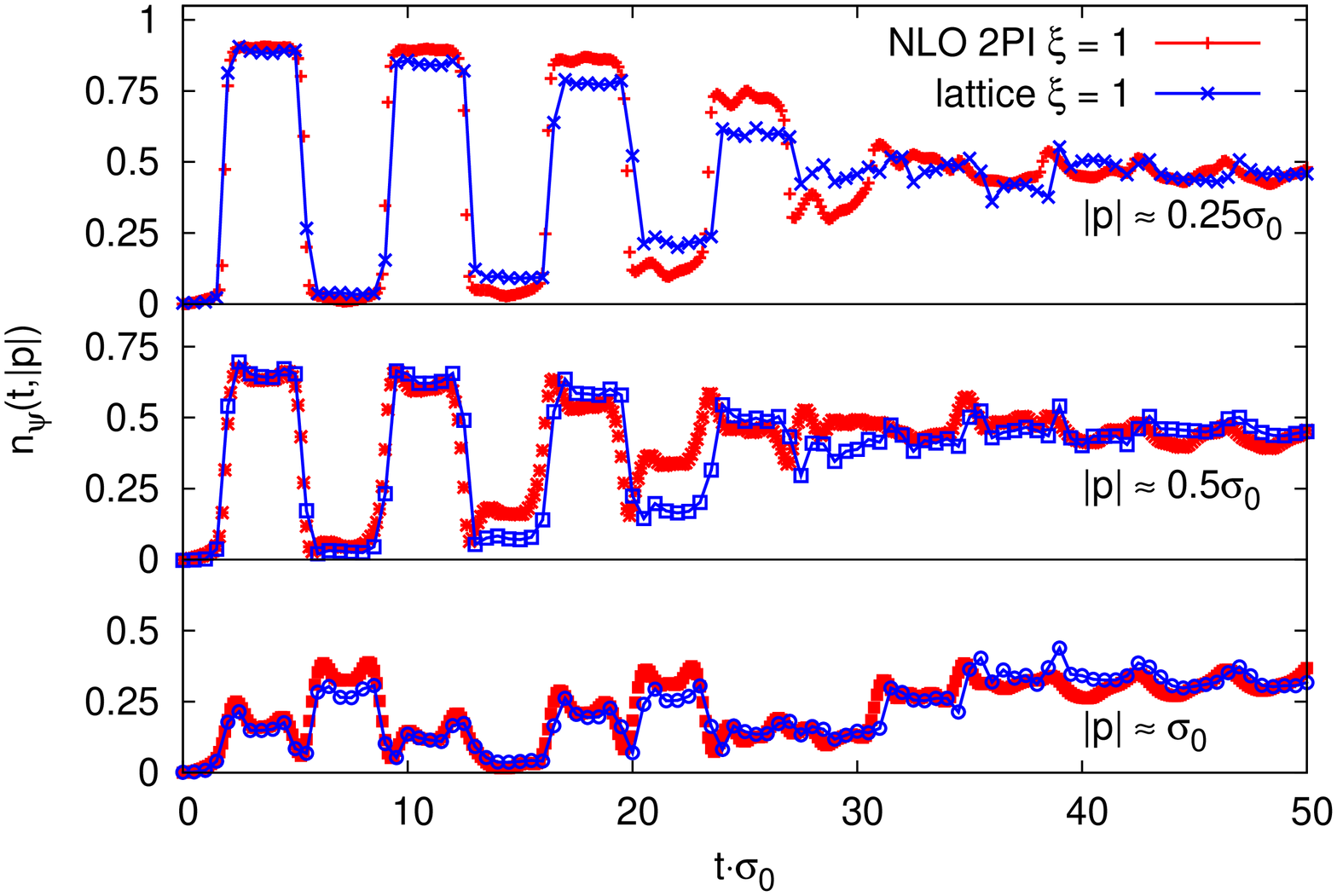}} &
      \resizebox{85mm}{!}{\includegraphics[width=\mywidth, angle=270]{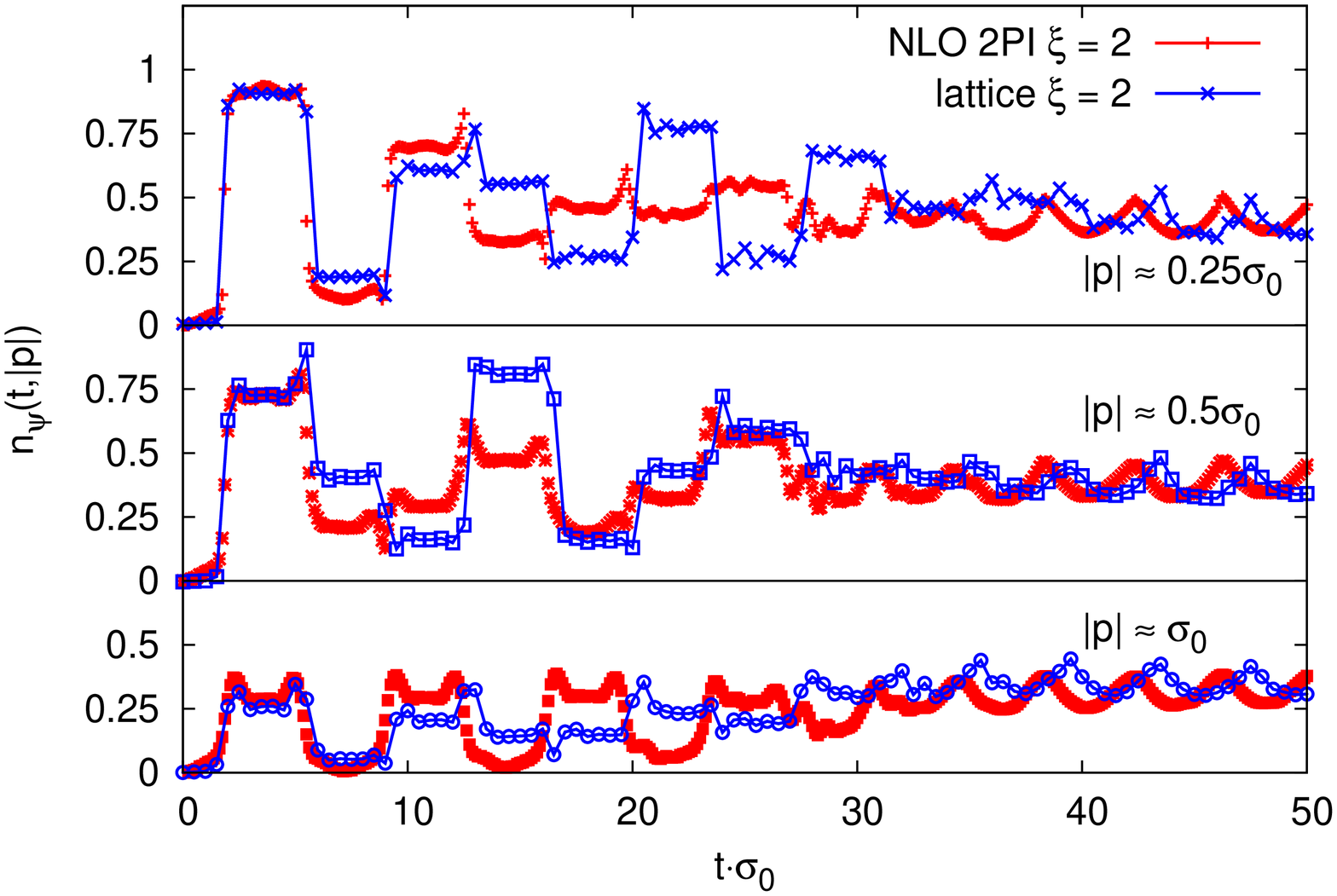}} \\
    \end{tabular}
    \caption{Time evolution of fermion occupation numbers for three different momenta at effective coupling strengths in the range $\xi=0.1 - 2.0$.}
    \label{fig:2PI_momenta_xi01-2}
  \end{center}
\end{figure*}

In Fig.~\ref{fig:2PI_momenta_xi01-2} we plot the fermion number $n_\psi(t,|{\bf p}|)$ as a function of time for three different momentum modes $|{\bf p}|= 0.25 \sigma_0$, $0.5 \sigma_0$ and $\sigma_0$. Since we expect our 2PI effective action approximation for the quantum evolution to break down at strong coupling, we give the results for different values of the effective coupling $\xi= g^2/\lambda$ and compare them to the respective lattice simulation results. One observes from Fig.~\ref{fig:2PI_momenta_xi01-2} that for $\xi=0.1$ the agreement between quantum and lattice approach is almost perfect. It worsens with increasing coupling as expected, but even at $\xi=1.0$ infrared modes seem to be quiet accurately reproduced. However, at $\xi=2$ the coupling expansion seems to finally break down.
\begin{figure}[h]
\begin{center}
\scalebox{0.7}{\epsfig{file=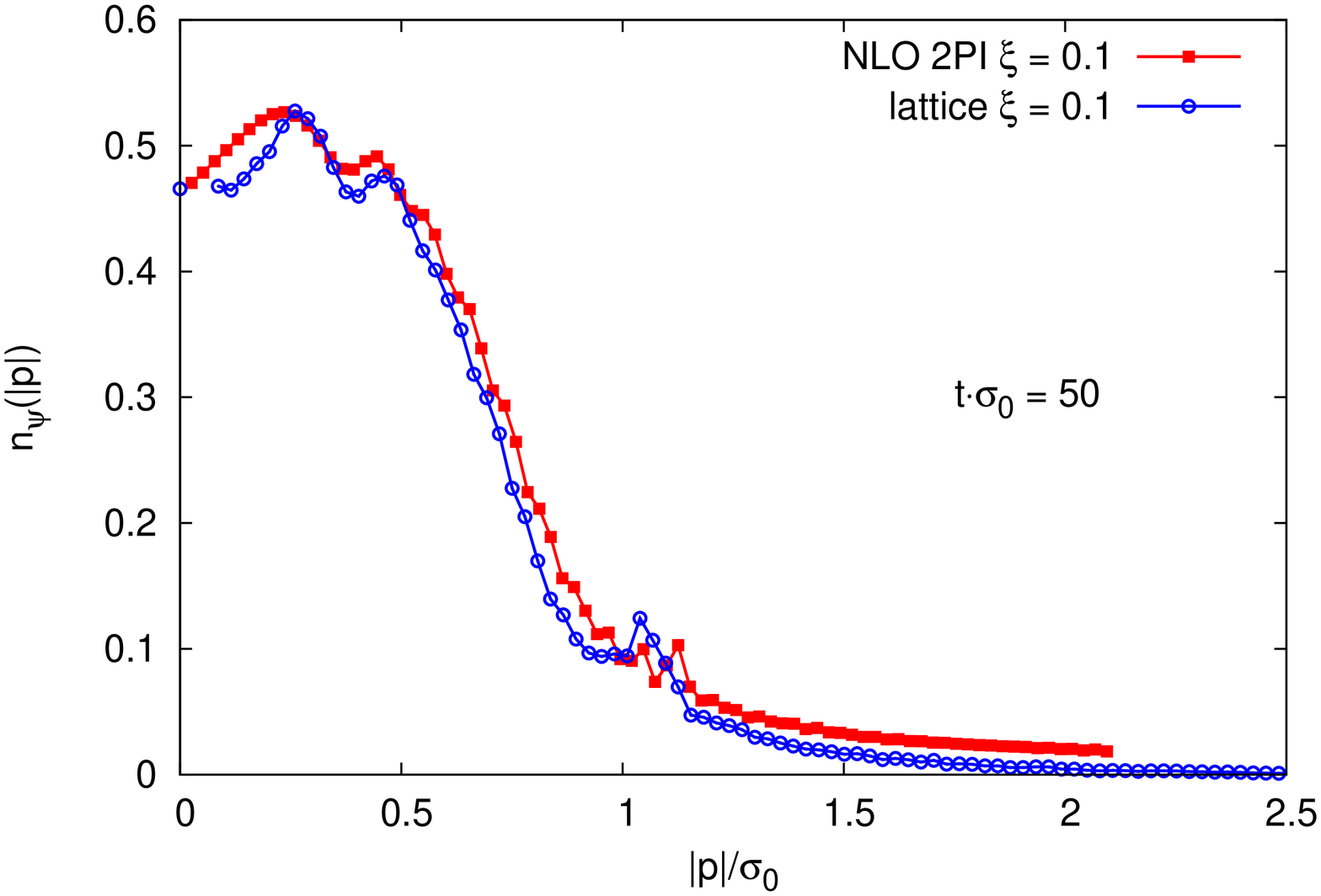, width=\mywidth, angle=270}}
\caption{Comparison between fermion spectra at $\xi=0.1$.}
\label{fig:2PI_spectra_xi01}
\end{center}
\end{figure}
\begin{figure}[h]
\begin{center}
\scalebox{0.7}{\epsfig{file=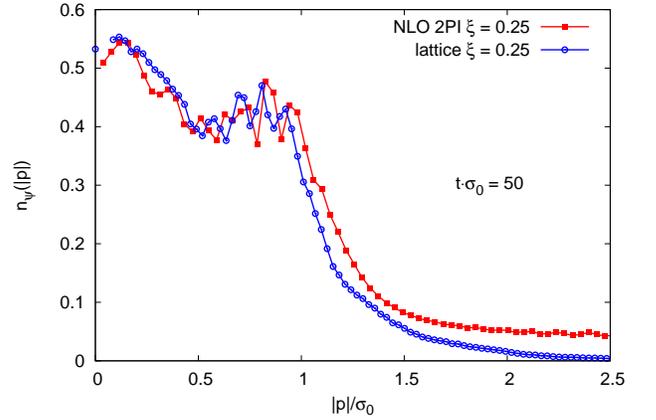, width=\mywidth, angle=270}}
\caption{Comparison between fermion spectra at $\xi=0.25$.}
\label{fig:2PI_spectra_xi025}
\end{center}
\end{figure}

In Figs.~\ref{fig:2PI_spectra_xi01} and \ref{fig:2PI_spectra_xi025} we present the full spectrum at fixed time $t \sigma_0 = 50$ for two different couplings. We observe a rather good agreement between both methods, reproducing characteristic features in the shape of the distribution. However, there is a clear discrepancy in the high-momentum part of the spectrum building up for the larger $\xi=0.25$. 
In general, we find good agreement between both methods at sufficiently small $\xi$ as expected while the agreement worsens for larger $\xi$. The level of agreement for small $\xi$ is also remarkable since the comparison involves two very different procedures: The results from the quantum 2PI effective action approach are obtained from a single run of the time evolution equations for correlation functions, while the lattice results are computed from a statistical average of many different runs of the corresponding lattice evolution equations. Already because of the Wilson term for the lattice description, it is rather difficult to get precisely the same initial conditions realized in both cases.  

These results confirm that possible dicretization and statistical errors on the lattice are under control. On the other hand, they show that for not too strong coupling a loop approximation beyond lowest order is sufficient to describe fermion production accurately. This has to be confronted with standard semi-classical (LO) descriptions of fermion production, which employ the solution of a Dirac equation in the presence of a time-dependent but spatially homogeneous background field neglecting fluctuations \cite{LOpapers}. In this case, the evolution equation for the fermion statistical two-point function reads
\begin{equation}
\left[ i \gamma^\mu \partial_{x,\mu} - \frac{g}{2} \phi(t) \right] F_\psi(x,y) = 0 \, .
\label{eq:LOFpsi}
\end{equation}
In contrast, our employed NLO 2PI approximation includes one-loop self-energy corrections to this equation. 
From the point of view of the lattice approach, a crucial difference concerns the spatial dependence of the fluctuating fields appearing in (\ref{Dirac_equations}). Sampling with respect to these fluctuations leads to the generation of loop corrections for ensemble averages, which are missing in (\ref{eq:LOFpsi}).
In the next section we will present numerical evidence that a semi-classical approximation fails to describe the dynamics and may only be applied for very short times of the initial evolution.

\section{Fermion production from parametric resonance}\label{parametric}

\begin{figure}[t!]
\begin{center}                                     
\scalebox{0.7}{\epsfig{file=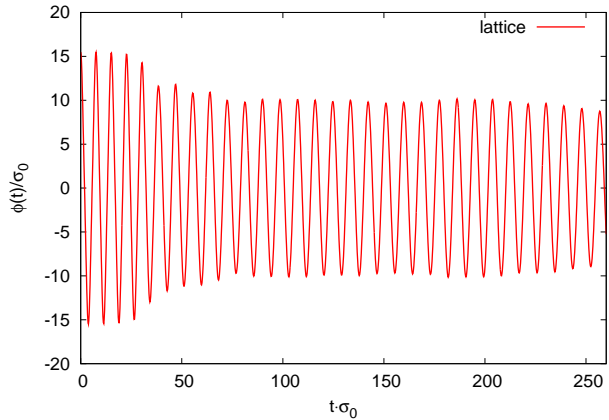, width=\mywidth, angle=270}}
\caption{Time evolution of the macrsocopic field.}
\label{fig:phiintime}
\end{center}
\end{figure}

\begin{figure}[t!]
\begin{center}
\scalebox{0.7}{\epsfig{file=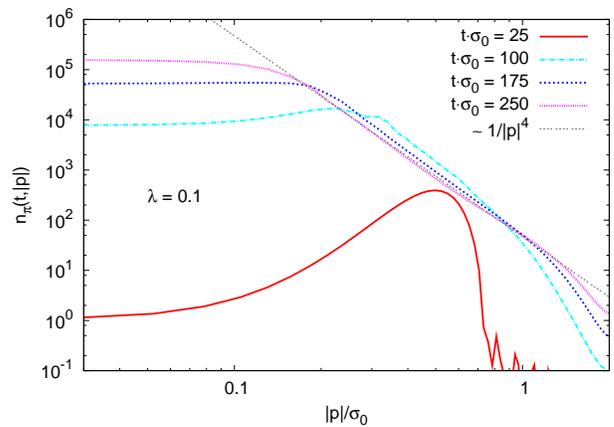, width=\mywidth, angle=270}}
\caption{Spectra of transverse scalar occupation numbers at different times.}
\label{fig:boson_spectra}
\end{center}
\end{figure}

The initial conditions described in section \ref{model} lead to the well-known phenomenon of parametric resonance in the scalar sector which we summarize~\cite{Kofman:1994rk,Khlebnikov:1996mc,Berges:2002cz}: Small initial quantum fluctuations grow exponentially in time. At early times this growth occurs in a compact momentum range with $\vec{p}^2 \leq \sigma_0^2/2$. As time proceeds, the exponentially growing modes induce non-linear behavior and secondary instabilities for 
higher momentum modes with even faster growth rates occur. As a consequence, there is a fast rise in the occupation numbers $n_{\sigma,\pi}(t,|{\bf p}|)$ for a broad momentum range. Figs.~\ref{fig:phiintime} and \ref{fig:boson_spectra} show the behavior of the field $\phi(t)$ and the occupancies of transverse modes $n_\pi(t,|{\bf p}|)$, respectively. The rapidly oscillating field decreases its amplitude with time, while the occupancies grow. After the fast initial growth period of occupation numbers, the time evolution of the now highly occupied scalar field modes slows down considerably, and can be described in terms of turbulent flows of energy and particle number \cite{StrongWeakTurbulence,BoseEinstein,NowakSextyGasenzer,Gasenzer:2013era}. The subsequent evolution becomes self-similar and the corresponding power-law behavior with $n_\pi(|{\bf p}|) \sim 1/|{\bf p}|^4$ is clearly visible from Fig.~\ref{fig:boson_spectra}. 


\begin{figure}[t!]
\begin{center}
\scalebox{0.7}{\epsfig{file=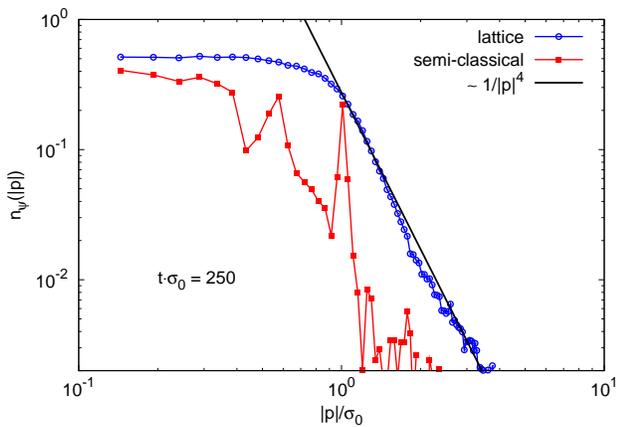, width=\mywidth, angle=270}}
\caption{Lattice simulation results for the occupation number distribution of weakly coupled fermions with $\xi = 0.1$ are compared to the standard semi-classical approximation.}
\label{fig:weak_powerlaw}
\end{center}
\end{figure}

In the following we will analyse the behavior of the fermions, which is the main topic of our work. To this end, it is useful to compare our lattice simulation results with standard semi-classical approximations based on equation (\ref{eq:LOFpsi}). For this comparison we distinguish the weak-coupling regime, where $\xi= q^2/\lambda \ll 1$, and the strongly coupled case with $\xi$ of order one. Fig.~\ref{fig:weak_powerlaw} shows the fermion occupation number distribution for $\xi = 0.1$ at the time $t \sigma_0 = 250$ after the initial instability has ceased. The lattice simulation results (circles) show a low-momentum range for $|\vec{p}| \leq \sigma_0$, where the distribution is rather flat. For higher momenta one observes a power-law behavior whose exponent agrees well with the scaling exponent found for the bosonic occupancies as shown in Fig.~\ref{fig:boson_spectra}. While bosons can support a $1/|{\bf p}|^4$ dependence also at low momenta, of course the fermion number distribution has to level off in the infrared because of the Pauli exclusion principle \cite{Berges:2002wr,PrevFerm}.
For this weak-coupling case we observe corresponding results also in the quantum theory based on the 2PI effective action at NLO, in accordance with the discussion of Sec.~\ref{2PI}. However, the lattice results including flucutuations clearly show significant differences to the semi-classical approximation results (squares), which neglect all fluctuations. In fact, the observed differences are so pronounced only after the bosonic fields become highly populated. This enhancement of fluctuations, which is missing in standard semi-classical approximations, will be discussed in detail below.

\begin{figure}[t!]
\begin{center}
\scalebox{0.7}{\epsfig{file=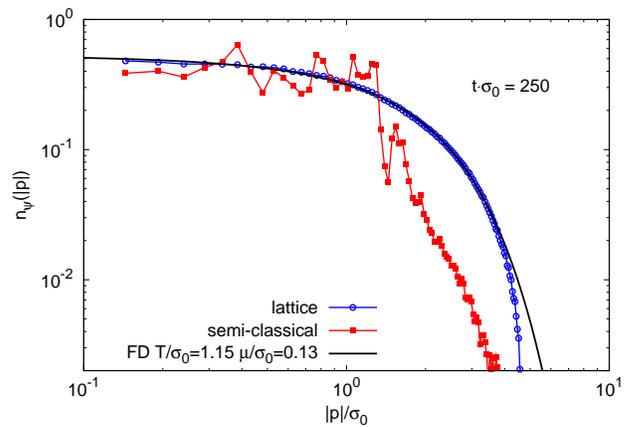, width=\mywidth, angle=270}}
\caption{Lattice simulation results for the occupation number distribution of strongly coupled fermions with $\xi = 1$ are compared to the semi-classical approximation. Shown is also a Fermi-Dirac distribution with time-dependent temperature and chemical potential parameters.}
\label{fig:strong_FD}
\end{center}
\end{figure}

\begin{figure}[t!]
\begin{center}
\scalebox{0.7}{\epsfig{file=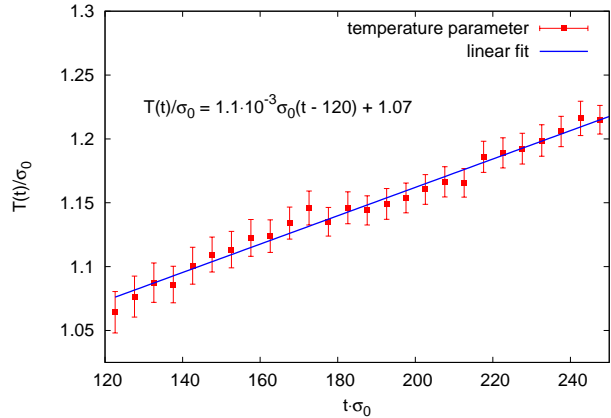, width=\mywidth, angle=270}}
\caption{Time-dependent temperature parameter as a function of time.}
\label{fig:T_av_time}
\end{center}
\end{figure}

A similar snapshot of occupation number spectra for the strongly coupled case ($\xi = 1$), shown in Fig.~\ref{fig:strong_FD}, reveals a rather different picture. Here, the lattice simulation results exhibit a distribution without any powerlaw behavior. Remarkably, the distribution can be nicely fitted to a Fermi-Dirac distribution with time-dependent temperature and chemical potential parameters.  At the time $t \sigma_0 = 250$ employed for Fig.~\ref{fig:strong_FD} they are $T/\sigma_0 = 1.15$ and $\mu/\sigma_0=0.13$. The similarity to the Fermi-Dirac distribution is non-trivial at this stage, since the bosons are still far from equilibrium showing the characteristic $\sim 1/|{\bf p}|^4$ behavior in the infrared. The Fermi-Dirac distribution also requires the specification of a dispersion relation or $\omega_p$, and we approximate it here by the free dispersion relation for massless fermions with a pseudoscalar Wilson term as discussed in (\ref{dispersion_PS}). It should be emphasized that the total charge in our simulations is zero such that the number of particles and antiparticles are equal. As a consequence, the chemical potential vanishes for true thermal equilibrium. Here we find that $\mu(t)$ is oscillating, which is not surprising in view of the oscillating Yukawa fermion mass term at this stage, and its absolut value turns out to be much smaller than the temperture for all considered times. The time-dependence of the fitted temperature parameter is shown in Fig.~\ref{fig:T_av_time}. It is seen to approximately rise linearly in time, after performing an average over periods of $\triangle t = 5/\sigma_0$ in order to smooth oscillations due to $\phi(t)$.
To get a simple estimate of how much this time-dependent temperature parameter deviates from the value of the equilibrium termperature, we compute the temperature of a corresponding gas of non-interacting massless bosons and fermions having a continuum dispersion relation:
\begin{equation}\label{eqn:temperature}
T_{\rm eq} = \sigma_0\left(\frac{45N_s}{\pi^2\left(N_s+\frac{7}{2}N_f\right)\lambda}\right)^{\frac{1}{4}} \simeq 2.02\sigma_0 \, .
\end{equation}
Here we used the initial energy density, which for parametric resonance is given in terms $\phi(t=0)$ and the self-coupling $\lambda$ (here $\lambda=0.1$). This estimate indicates that the observed time-dependent temperature parameter is still far away from the asymptotic equilibrium value. The observed linear rise of the temperature parameter would lead to the above estimate for the equilibrium temperature after a time $t_{\rm eq} \simeq 10^3/\sigma_0$.

\begin{figure}[t!]
\begin{center}                                     
\scalebox{0.7}{\epsfig{file=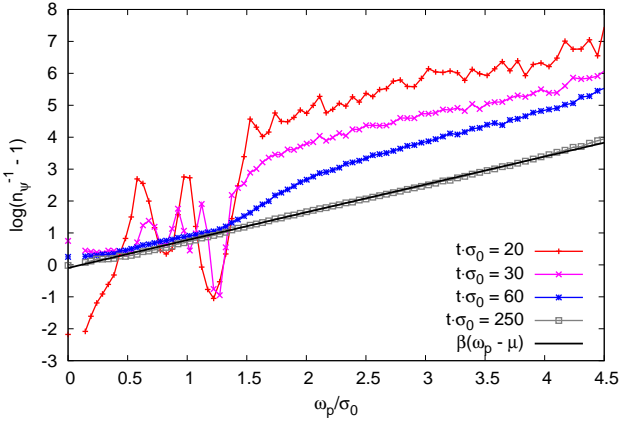, width=\mywidth, angle=270}}
\caption{Lattice simulation results for the 'inverse slope parameter' ${\rm ln}\left(n_{\psi}^{-1}-1\right)$ at four different times. For comparison the solid line corresponds to a Fermi-Dirac distribution with fitted temperature and chemical potential at $t\sigma_0 = 250$.}
\label{fig:strong_lognp}
\end{center}
\end{figure}

To study in more detail the similarities and deviations from a Fermi-Dirac distribution, it is instructive
to consider the 'inverse slope parameter' ${\rm ln}\left(n_{\psi}^{-1}-1\right)$. Fig.~\ref{fig:strong_lognp} shows this quantity for $\xi = 1$ as a function of $\omega_p$ at four different times. For a thermal equilibrium distribution it would be a time-independent straight line. For a vanishing chemical potential in thermal equilibrium this line would go through the origin.
From the figure one observes that rather quickly an approximately stable inverse slope is established for lower momenta around $|{\bf p}| \lesssim 1.5 \omega_p/\sigma_0$. This is due to the fact that already at early times many low-momentum bosonic quanta are occupied, making it possible for fermions to scatter off them and redistribute momentum and energy. In contrast, there are almost no high-momentum bosons present at early times, preventing a more efficient transfer of energy and particles to the UV. Around $t \sigma_0 = 250$ one observes again the high level of agreement with a Fermi-Dirac distribution with fitted temperature and chemical potential parameter at that time. 

\section{Understanding amplified fermion production for large bosonic occupancies}\label{kinetic}

The above lattice simulation results for fermion production from parametric resonance revealed a dramatic difference compared to standard semi-classical estimates. The latter includes for instance fermion decay from a homogeneous background field, but neglects fluctuations or scattering effects. Scattering processes can become strongly enhanced if the participating modes are highly occupied. Correspondingly, the observed differences between lattice simulations and semi-classical treatment are pronounced once the bosons become strongly occupied.

In the following we analyse the impact of scattering processes on the dynamics in more detail. To this end, we consider the NLO approximation for the quantum 2PI effective action of Sec.~\ref{2PI} in the weak-coupling regime with $\xi = 0.1$. As discussed above, in this regime the NLO 2PI approximation is found to accurately reproduce the full lattice simulation result. In particular, we will use this approximation to derive kinetic equations that explain the relevant underlying processes. The power counting will be based on an expansion in the coupling $g$ in the presence of large occupancies with parametric dependence $n_{\sigma,\pi}(t,|{\bf p}|) \sim 1/\lambda$ for $\lambda \ll 1$ and $g^2/\lambda \ll 1$. We also emphasize that the observation of an amplified fermion production in the presence of large bosonic occupancies is not specific to the phenomenon of parametric resonance, though we will continue to consider this example. For the main points of the following analysis one could equally well consider other nonequilibrium instabilities leading to -- or even directly starting from -- large occupancies. 

\begin{figure}[t!]
\begin{center}                                     
\scalebox{0.7}{\epsfig{file=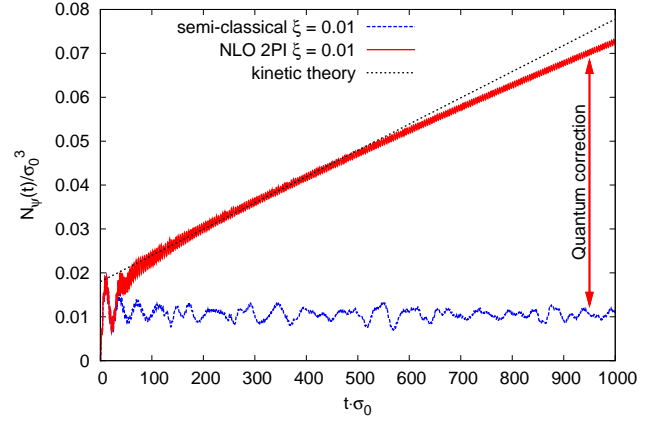, width=\mywidth, angle=270}}
\caption{Total number of produced fermions at order $g$ (semi-classical approximation, dashed line) and order $g^2$ (NLO 2PI, solid line) for the weak-coupling regime with $\xi = g^2/\lambda = 0.01$. Shown is also an analytic estimate for the production rate from kinetic theory (dotted line), which is shifted to account for the initial semi-classical fermion production until $t\sigma_0 \simeq 10$.}
\label{fig:Ntot}
\end{center}
\end{figure}

Counting powers of the coupling $g$, direct scattering appears at order $g^2$ according to Sec.~\ref{2PI}, whereas the semi-classical approximation based on equation (\ref{eq:LOFpsi}) is restricted to processes at order $g$. Fig.~\ref{fig:Ntot} compares the total number of produced fermions at order $g^2$ (NLO 2PI, solid line) to results from the semi-classical approximation (dashed line). In the figure we also give an analytic estimate for the production rate from kinetic theory (dotted line), which will be explained below. While for very short times the order $g$ and $g^2$ results results agree rather well, after the end of the parameteric resonance regime, i.e.\ when the occupancies become large, the order $g^2$ corrections are seen to dominate the fermion production by far. Apparently, highly occupied bosons act as a very efficient amplifier for genuine quantum corrections to the fermion dynamics. 

This phenomenon can be understood from the NLO approximation of the 2PI effective action described in Sec.~\ref{2PI}. In order to make analytic progress, we consider in addition a standard gradient expansion
to lowest order in derivatives following Ref.~\cite{Berges:2005md}. It employs for two-point functions, such as $F_{\phi}(x,y)$ given by (\ref{eq:Fphi}) and $F_{\psi}(x,y)$ defined in (\ref{eq:Fpsi}), the introduction of relative coordinates $x-y$ and center coordinates $(x+y)/2$. For the considered spatially homogeneous systems, a Fourier transformation with respect to the relative coordinates leads to $F_{\sigma,\pi}(t,k)$ and $F_{\psi}(t,k)$ with four-momentum $k$ and the time coordinate $t \equiv (x^0+y^0)/2$.
The general form of the equation for the boson two-point function to lowest order in derivatives with respect to the center coordinate is
\begin{equation}\label{eqn:Boltzmann1}
2k^0 \partial_t F_{\sigma,\pi}(t,k) = \Sigma_{\rho}(t,k) F_{\sigma,\pi}(t,k) - \Sigma_{F}(t,k) \rho_{\sigma,\pi}(t,k) \, ,
\end{equation}
where the 'collision term' on the right hand side encodes the 'gain' minus 'loss' structure in terms of the statistical ($\Sigma_{F}$) and spectral ($\Sigma_{\rho}$) parts of the self-energy. At order $g^2$ the spectral and statistical components of the self-energy displayed on the right of Fig.~\ref{fig:renorm_diagrams} read
\begin{eqnarray}
\nonumber\Sigma_{F}(t,k)&=&\frac{g^2}{2}\int\frac{d^4 p}{(2 \pi)^4}Tr [ F_{\psi}(t,k+p)F_{\psi}(t,p)\\
&&-\frac{1}{4}\rho_{\psi}(t,k+p)\rho_{\psi}(t,p)] \, ,
\\
\nonumber\Sigma_{\rho}(t,k)&=&\frac{g^2}{2}\int\frac{d^4 p}{(2 \pi)^4}Tr [ F_{\psi}(t,k+p)\rho_{\psi}(t,p)\\
&&-\rho_{\psi}(t,k+p)F_{\psi}(t,p)] \, .
\end{eqnarray}
The physical content of these expressions can be further clarified by setting the fermion spectral function $\rho_{\psi}(t,k)$ on shell and introducing occupation numbers $n_{\phi}(t,k)$ and $n_{\psi}(t,k)$ for bosons and fermions, respectively, with 
\begin{eqnarray}
\nonumber F_{\sigma,\pi}(t,k)&=&\left[ \frac{1}{2}+n_{\phi}(t,k) \right]\rho_{\sigma,\pi}(t,k) \, ,
\\
F_{\psi}(t,k)&=&\left[ \frac{1}{2}-n_{\psi}(t,k) \right]\rho_{\psi}(t,k) \, .
\end{eqnarray}
The (anti-)symmetry of the (spectral) statistical correlation functions translate into
\begin{equation}
n_{\phi}(t,-k) = -\left[ n_{\phi}(t,k)+1  \right] \, , \, \, n_{\psi}(t,-k) = -\left[ n_{\psi}(t,k)-1  \right] \, .
\end{equation}
The time-independent on-shell spectral functions are given by 
\begin{eqnarray}
\rho_{\sigma,\pi}(k)=2\pi {\rm sgn}(k^0) \delta(k^2-m^2) \, ,
\\
\rho_{\psi}(k)=2\pi k_{\mu}\gamma^{\mu} {\rm sgn}(k^0) \delta(k^2) \, ,
\end{eqnarray}
for the considered case of massless fermions. After performing traces in spinor and flavour space as well as some of the integrals and projecting onto positive frequencies we arrive at 
\begin{eqnarray}
\partial_t n_{\phi}(t,{\bf k}) &=& \pi g^2 \!\int \! \frac{d^3 p}{(2\pi)^3} \int \! d^3 q \delta\left({\bf k} - {\bf p} - {\bf q} \right) \delta \left( \omega_k - |{\bf p}| - |{\bf q}| \right)  
\nonumber\\
&& \frac{1}{\omega_{k}} \left( 1 - \frac{{\bf p}{\bf q}}{|{\bf p}||{\bf q}|} \right) 
\left[ \left(n_{\phi}(t,{\bf k})+1\right)n_{\psi}(t,{\bf p})n_{\psi}(t,{\bf q}) \right.
\nonumber\\
&& \left. - n_{\phi}(t,{\bf k})\left(n_{\psi}(t,{\bf p})-1\right)\left(n_{\psi}(t,{\bf q})-1\right) \right] \, .
\label{eqn:Boltzmann2}
\end{eqnarray}
Here $\omega_{k}=\sqrt{k^2+m^2}$ is the free bosonic dispersion relation. Along the same lines one can obtain the corresponding kinetic equation for the fermion occupation number,
\begin{eqnarray}
\partial_t n_{\psi}(t,{\bf k}) &=& \pi g^2 \!\int \! \frac{d^3 p}{(2\pi)^3} \int \! d^3 q \delta\left({\bf k} + {\bf p} - {\bf q} \right) \delta \left( |{\bf k}| + |{\bf p}| - \omega_q \right)  
\nonumber\\
&& \!\!\!\!\!\!\!\!\!\!\!\!\frac{1}{\omega_{q}} \left( 1 - \frac{{\bf k}{\bf p}}{|{\bf k}||{\bf p}|} \right) 
\left[ \left(n_{\psi}(t,{\bf k})-1\right)\left(n_{\psi}(t,{\bf p})-1\right)n_{\phi}(t,{\bf q}) \right.
\nonumber\\
&& \left. \!\!\!\!\!\!\!\!\!\!\!\! - n_{\psi}(t,{\bf k})\left(n_{\psi}(t,{\bf p})-1\right)\left(n_{\phi}(t,{\bf q})+1\right) \right] \, .
\label{eqn:Boltzmann_psi}
\end{eqnarray}
From these expressions we observe that 
\begin{equation}
\partial_t \left( N_\phi(t) + N_\psi(t) \right) = \int \! \frac{d^3 k}{(2\pi)^3} \partial_t \left( n_{\phi}(t,{\bf k}) + n_{\psi}(t,{\bf k}) \right) = 0
\label{eq:cons}
\end{equation}
reflecting total number conservation of bosons and fermions, $N_\phi + N_\psi$, in this approximation. Total number changing processes would enter the kinetic description at higher order in $g$. Though they are crucial for the approach to thermal equilibrium at late times~\cite{Berges:2002wr,Lindner:2007am}, these processes turn out not to be important for the time of enhanced fermion production in the weak-coupling regime.

It is instructive to consider (\ref{eqn:Boltzmann2}) for $n_\psi = 0$, which is approximately realized at sufficiently early times. The equation can then be written as
\begin{equation}
\partial_t n_{\phi}(t,{\bf k}) \simeq - \Gamma_{\phi \rightarrow \psi \bar{\psi}}({\bf k})\, n_{\phi}(t,{\bf k})
\end{equation}
with
\begin{eqnarray} 
\Gamma_{\phi \rightarrow \psi \bar{\psi}}({\bf k}) &=& \pi g^2 \!\int \! \frac{d^3 p}{(2\pi)^3} \int \! d^3 q\, \delta\left({\bf k} - {\bf p} - {\bf q} \right)   
\nonumber\\
&& \delta \left( \omega_k - |{\bf p}| - |{\bf q}| \right) \frac{1}{\omega_{k}} \left( 1 - \frac{{\bf p}{\bf q}}{|{\bf p}||{\bf q}|} \right) \, . 
\end{eqnarray}
For ${\bf k} = 0$ one obtains the standard vacuum decay rate for the production of massless fermions with momenta $\pm m/2$, i.e.~$\Gamma_{\phi \rightarrow \psi \bar{\psi}}(0) = g^2 m/(8 \pi)$. Taking the number conservation (\ref{eq:cons}) into account, we can write for the change in the total fermion number
\begin{equation}
\partial_t N_\psi(t) \simeq \int \frac{d^3 k}{(2\pi)^3} \Gamma_{\phi \rightarrow \psi \bar{\psi}}({\bf k})\, n_\phi(t,{\bf k}) \, .
\end{equation}
To get a parametric estimate for the right hand side, we may approximate $n_\phi(t,{\bf k}) \simeq  \Theta(|{\bf k}| - \sigma_0)/\lambda$ around the time 
after the parametric resonance regime ends. The crucial ingredient is here the enhancement by a factor of $1/\lambda$, which for the considered weak-coupling case encodes the amplification of $\partial_t N_\psi$ from being order $g^2$ to order $g^2/\lambda \equiv \xi$ for parametrically large Bose occupancies. As a consequence, one expects an approximately linear rise in the total fermion number with slope proportional to $\xi$ as shown in Fig.~\ref{fig:Ntot} for $\xi = 0.01$.
Because of the non-zero fermion occupation numbers building up with time, this linear rise is diminished by the Pauli suppression due to the presence of already produced fermions. We can also use (\ref{eqn:Boltzmann2}) to estimate the magnitude of the backreaction of fermions onto the bosonic sector. To this end, we compare the bosonic 'gain term' $\sim n_{\psi}(t,{\bf p})n_{\psi}(t,{\bf q})$ to the 'loss term' $\sim -n_{\phi}(t,{\bf k})(1-n_{\psi}(t,{\bf q})- n_{\psi}(t,{\bf k})$. The latter is enhanced by the macroscopic occupation of scalars while the former is strictly $\leq 1$ and we find it to be even $\leq 1/4$ at later times during our simulations. The minor role of fermionic backreaction in the weak-coupling regime agrees well with our findings from the full simulation data.

\begin{figure*}[t!]
  \begin{center}
    \begin{tabular}{cc}
      \resizebox{70mm}{!}{\includegraphics[width=\mywidth, angle=270]{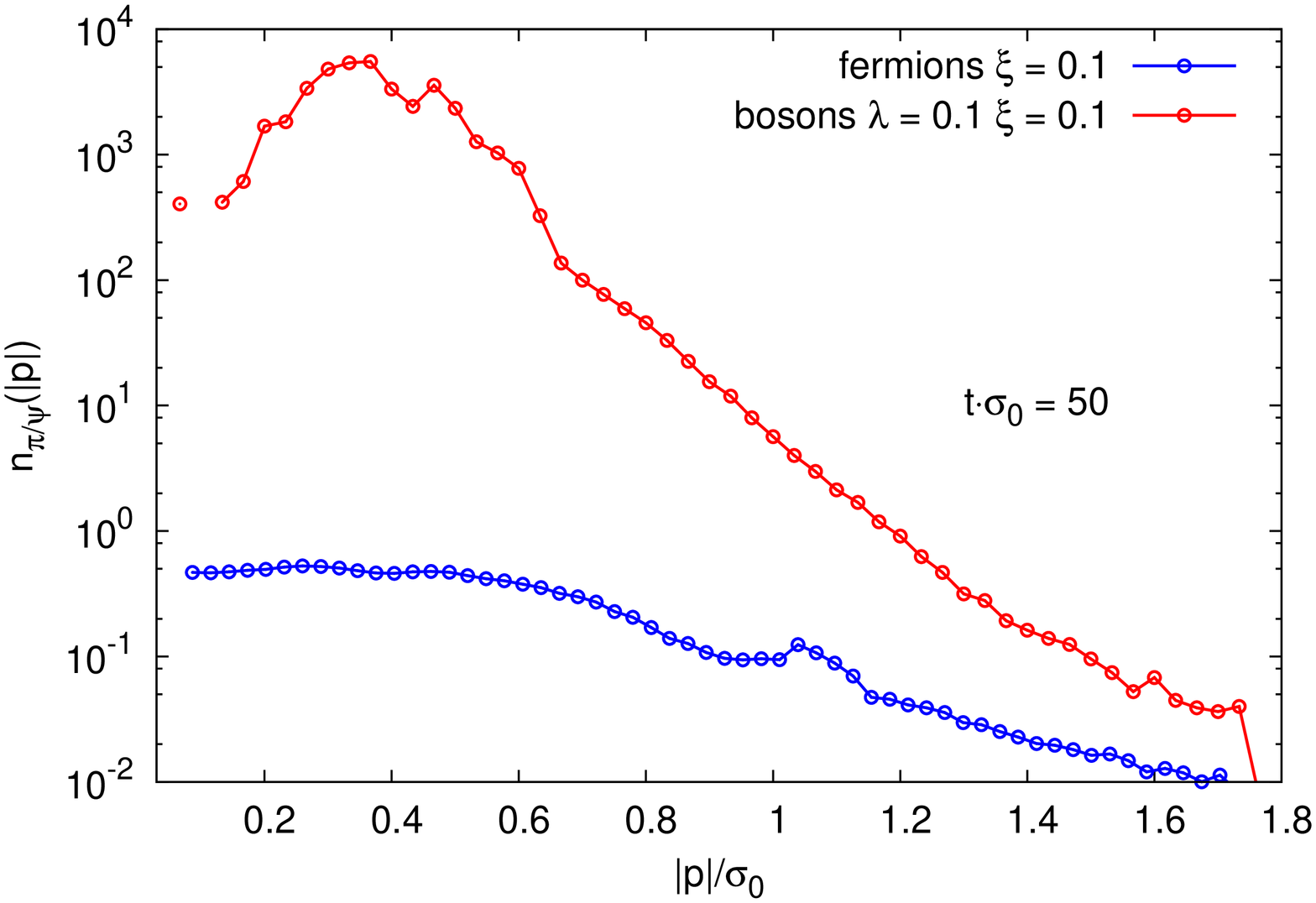}} &
      \resizebox{70mm}{!}{\includegraphics[width=\mywidth, angle=270]{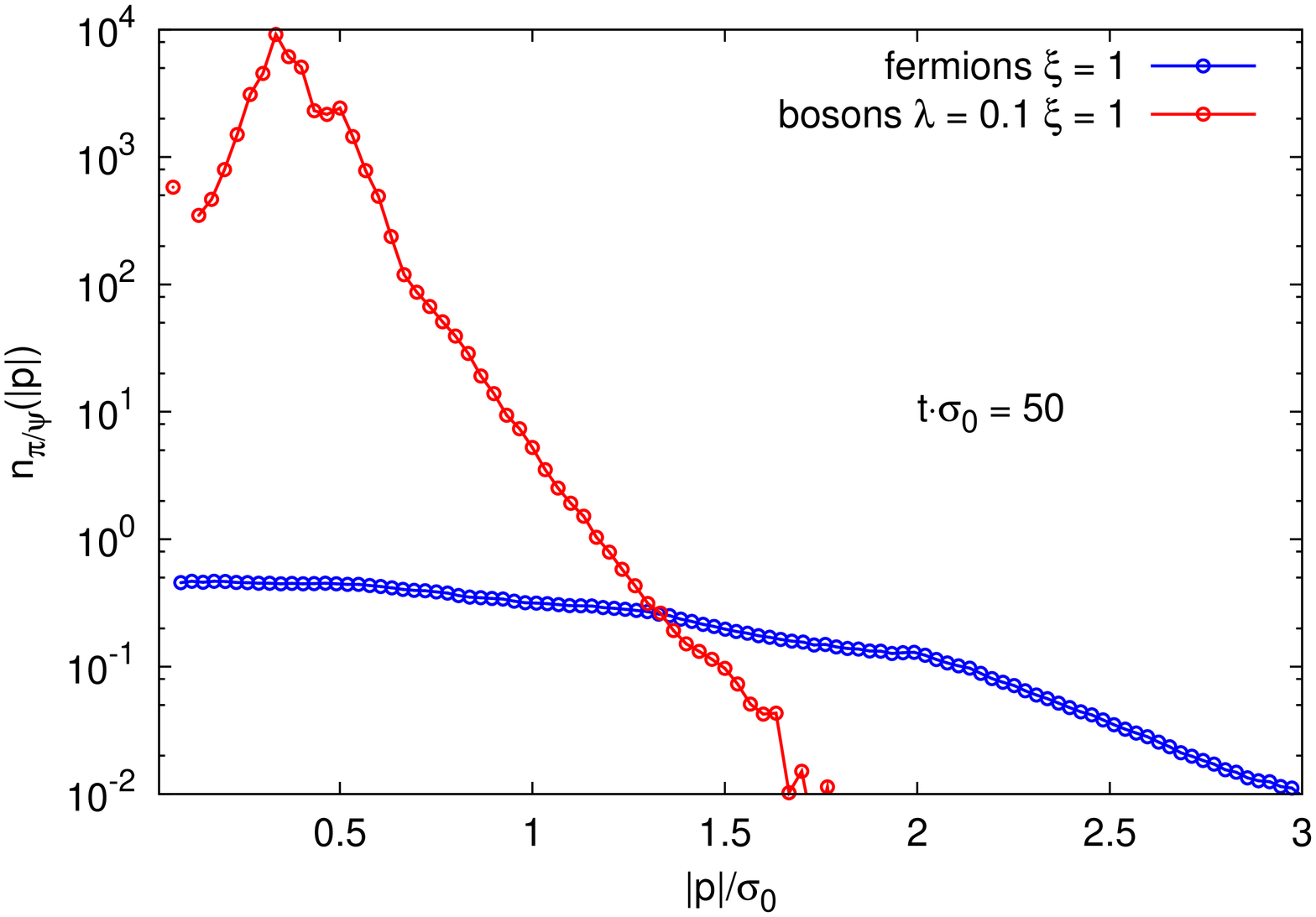}} \\
    \end{tabular}
    \caption{Lattice simulation results for occupation number distributions of fermions and bosons at $t \sigma_0 = 50$ for $\xi=0.1$ and $\xi=1$.}
    \label{fig:UV_multiple_scattering}
  \end{center}
\end{figure*}

The above kinetic description provides a detailed understanding of the weak-coupling case with $\xi \ll 1$. It is expected to fail to describe the physics for strong couplings, where higher order processes are no longer suppressed. This is also what we find by comparing it to the nonperturbative lattice simulation results in accordance with the discussion of Sec.~{2PI}. It is a characteristic property of the above kinetic description that typical fermion and boson momenta are similar. In Fig.~\ref{fig:UV_multiple_scattering} on the left we show the occupancies of bosons and produced fermions for $\xi = 0.1$, where the highest occupied momenta agree rather well. The right graph shows the same quantities for $\xi = 1$, where one observes the tendency to occupy higher maximum momenta for fermions than for bosons. Even though the fermion occupancy per mode is limited by the exclusion pronciple, they quickly populate higher momentum modes in the nonperturbative regime. This contribution becomes essential for strong enough coupling since the high-momentum part carries most of the energy. 

\section{Conclusions}\label{Conclusions}

In this work we have studied nonequilibrium production of fermions  
from parametric resonance in 3 + 1 dimensions for a generic linear  
sigma model. As our main result, we confirmed the dramatic  
amplification of fermion production in the presence of highly occupied  
bosons that was first pointed out in Ref.~\cite{BergesGelfandPruschke}  
and extended the results to the strong coupling regime.

We compared different real-time techniques -- lattice simulations with  
male/female fermions, mode functions approach and quantum 2PI  
effective action with its associated kinetic theory -- and discussed  
their range of applicability. It turned out that the efficient  
male/female lattice approach accurately converges to the exact mode  
functions result for the available lattice sizes. The study shows the  
strength of the male/female method to address physical questiones for  
large volumes, something where the mode function approach becomes  
computationally intractable. For weak couplings we found that the  
lattice simulation results agree well with those obtained from the  
quantum 2PI effective action, emphasizing the ability of the lattice  
approach to describe genuine quantum phenomena.

Applying an improved lattice discretization with a pseudoscalar Wilson  
term, we have been able to accurately resolve the high-momentum  
behavior of particle number distributions. For weak couplings this  
revealed a power-law behavior above a characteristic momentum. For 
strongly coupled fermions, we  
found that a quasi-thermal Fermi-Dirac distribution is  
approached, with time-dependent temperature and chemical potential  
parameters. This happens while the bosons are still showing turbulent  
behavior far from equilibrium \cite{AartsSmitTemp}.

In the employed model the coupling to the fermions and the bosonic  
self-coupling can be separately chosen to represent the weak-coupling  
$(\xi \ll 1)$ and the strong-coupling regime $(\xi \gtrsim 1)$. This  
allowed us to validate the lattice simulation techniques in the  
weak-coupling regime by comparing it to alternative, quantum  
techniques. The strategy for future studies is to employ these  
nonperturbative techniques to theories, where $\xi \ll 1$ cannot be  
realized. An important class of such theories concerns non-Abelian  
gauge theories, where the bosonic self-coupling and the coupling to  
the fermion sector are given by the same coupling such that the  
relevant ratio is $\xi = 1$.

We thank Gert Aarts, Szabolcs Borsanyi, Jens Pruschke, and Thorsten  
Z\"oller for discussions. This work is supported by the DFG and  
D.~Gelfand thanks HGS-HIRe for FAIR for support.

\end{document}